%% file: manuscript.tex
\documentclass[lettersize,journal]{IEEEtran}
\usepackage{amsmath,amsfonts}
\usepackage{algorithm}
\usepackage{algpseudocode}
\usepackage{array}
\usepackage{textcomp}
\usepackage{stfloats}
\usepackage{url}
\usepackage{verbatim}
\usepackage{graphicx}
\usepackage{tabularx, booktabs}
\usepackage{multirow}
\usepackage{subcaption}
\def\BibTeX{{\rm B\kern-.05em{\sc i\kern-.025em b}\kern-.08em
    T\kern-.1667em\lower.7ex\hbox{E}\kern-.125emX}}
\usepackage{balance}
\usepackage[switch]{lineno}
\newcolumntype{Y}{>{\centering\arraybackslash}X}

\newcommand{\FDG}{[$^{18}$F]FDG }

\begin{document}
\begin{titlepage}
\centering
\Large
\vspace*{\fill}
{\bfseries Personalized MR-Informed Diffusion Models for 3D PET Image Reconstruction\par}
\vspace{2\baselineskip}
This work has been accepted for publication in \emph{IEEE Transactions on Radiation and Plasma Medical Sciences}.  
The author’s accepted manuscript is released under a CC-BY licence.  
For the Version of Record, see DOI: 10.1109/TRPMS.2025.3602262.
\vspace*{\fill}
\end{titlepage}

\title{Personalized MR-Informed Diffusion Models for 3D PET Image Reconstruction}

\author{George Webber,~\IEEEmembership{Student Member,~IEEE}, Alexander Hammers, Andrew P. King, Andrew J. Reader
\thanks{G. Webber (e-mail: george.webber@kcl.ac.uk), A. P. King and A. J. Reader are with the School of Biomedical Engineering and Imaging Sciences, King’s College London, UK. A. Hammers is with King’s College London and Guy's \& St Thomas' PET Centre.}
\thanks{G. Webber would like to acknowledge funding from the EPSRC Centre for Doctoral Training in Smart Medical Imaging [EP/S022104/1] and via a GSK studentship. This work was also supported in part by the Wellcome/EPSRC Centre for Medical Engineering [WT 203148/Z/16/Z], and in part by EPSRC grant number [EP/S032789/1]. This work involved human subjects or animals in its research. The author(s) confirm(s) that all human/animal subject research procedures and protocols are exempt from review board approval.}
\thanks{For the purposes of open access, the authors have applied a Creative Commons Attribution (CC BY) licence to any Accepted Author Manuscript version arising, in accordance with King’s College London’s Rights Retention policy.}
}
% The paper headers
\markboth{Transactions on Radiation and Plasma Sciences, Vol XX, No XX, Month Year}%
{Title here}

\maketitle

\begin{abstract}
Recent work has shown improved lesion detectability and flexibility to reconstruction hyperparameters (e.g. scanner geometry or dose level) when PET images are reconstructed by leveraging pre-trained diffusion models. Such methods train a diffusion model (without sinogram data) on high-quality, but still noisy, PET images. In this work, we propose a simple method for generating subject-specific PET images from a dataset of multi-subject PET-MR scans, synthesizing ``pseudo-PET" images by transforming between different patients' anatomy using image registration. The images we synthesize retain information from the subject's MR scan, leading to higher resolution and the retention of anatomical features compared to the original set of PET images. With simulated and real \FDG datasets, we show that pre-training a personalized diffusion model with subject-specific ``pseudo-PET" images improves reconstruction accuracy with low-count data. In particular, the method shows promise in combining information from a guidance MR scan without overly imposing anatomical features, demonstrating an improved trade-off between reconstructing PET-unique image features versus features present in both PET and MR. We believe this approach for generating and utilizing synthetic data has further applications to medical imaging tasks, particularly because patient-specific PET images can be generated without resorting to generative deep learning or large training datasets.
\end{abstract}

\begin{IEEEkeywords}
Score-based Generative Modeling, Image Reconstruction Algorithms, Positron Emission Tomography, Magnetic Resonance Imaging, Synthetic Data
\end{IEEEkeywords}

\section{Introduction}

Positron emission tomography (PET) is a widely-used medical imaging modality for visualizing and quantifying patients' metabolic and biochemical processes \cite{shukla_positron_2006}. PET scans require patient exposure to ionizing radiation, so radiotracer doses are kept low to reduce radiation-associated risks \cite{nievelstein_radiation_2012}. As a consequence, PET scans can suffer from limited spatial resolution and high levels of noise, particularly in comparison to other common imaging modalities such as magnetic resonance imaging (MR) and computed tomography (CT) \cite{schwaiger_petct_2005}.

A common approach to improve the spatial resolution of PET scans is to guide the PET image reconstruction process with an image acquired from an anatomical modality (i.e. MR or CT) \cite{tong_image_2010, bai_mr_2013}. In this way, one can complement the functional information acquired by PET with spatially-precise anatomic details. For this work, we focus on the case of brain imaging with simultaneously acquired PET and MR datasets.

Although reconstruction incorporating anatomical guidance can yield higher-resolution PET images, the key issue with such methodologies is how to resolve cases where structures in the true PET radiotracer distribution do not match structures in the MR guidance image \cite{bai_mr_2013}.
Assuming that a dataset featuring suitable mismatch examples is available, supervised deep learning could be leveraged to resolve mismatches \cite{corda-dincan_syn-net_2020}. However, for the case of abnormal or irregular PET lesions, this assumption rarely holds by definition.

Recent advances in score-based generative modeling \cite{chung_score-based_2022, chung_decomposed_2023} (a framework encompassing diffusion models) have enabled new state-of-the-art deep-learned PET reconstruction algorithms that require only high-quality images for training \cite{singh_score-based_2024, webber_likelihood_2025}. This class of deep-learning algorithm is more flexible than previous classes, as it enables us to train a model that is agnostic to scanner geometry and injected dose. These algorithms have been shown to result in superior lesion detectability relative to other unsupervised reconstruction algorithms (i.e. algorithms trained without access to PET sinogram data) \cite{singh_score-based_2024}. Although these models are usually trained with images reconstructed from full-count PET acquisitions \cite{singh_score-based_2024, webber_likelihood_2025}, any images can be used as training data. In the ideal case, a generative model would be trained on vast volumes of data drawn from the actual subject-specific probability distribution of possible ground truth images.

In the absence of such a data source, we propose a method that enables the generation of large numbers of diverse and subject-specific training images that have improved spatial resolution compared to conventional PET images. We achieve this by transforming other subjects' PET images into the target subject's anatomical frame-of-reference using MR-to-MR registration maps, leading to images we refer to as ``pseudo-PET". The prior information in the diffusion model is hence learned from transformed PET images that retain some MR anatomical features, and so may be viewed as a joint anatomical and functional prior. As a result, this method has particular potential for MR-guided PET reconstruction, as a softer form of guidance than methods utilizing MR data directly.

We perform PET image reconstruction with this learned prior using the PET-DDS (Decomposed Diffusion Sampling) algorithm \cite{singh_score-based_2024}. We evaluate our approach on 3D simulated \FDG data with out-of-distribution lesions as well as real \FDG data.

We find that our approach to integrating MR guidance with PET is less prescriptive than other methods, delivering lower quantification errors in areas of mismatch between PET and MR than other methods. Our approach also yields superior overall performance (as measured with a bias-variance trade-off assessment) to PET-DDS trained with ordinary PET (and MR) images.

This work makes the following contributions:
\begin{itemize}
    \item We propose a general method for synthesizing high-quality subject-specific synthetic PET images, dubbed "pseudo-PET", that avoids the use of deep generative models by leveraging MR-based registration to an anatomically defined space.
    \item We integrate our approach for synthesizing subject-specific synthetic PET data with a state-of-the-art approach for PET reconstruction (using deep learning with PET image data).
    \item We show in reconstruction from 2.5\% low-count simulated data that our approach yields reduced error in areas of PET-MR mismatch without compromising error in areas of PET-MR agreement.
    \item We show results for MR-guided PET reconstruction from real \FDG data, and report that the PET-DDS approach \cite{singh_score-based_2024} for MR-guidance via classifier-free guidance does not always deliver the improvements seen in simulations (when trained on limited datasets).
    \item We show that our approach for 3D real \FDG data MR-guided reconstruction from 2.5\% low-count data yields improved bias-variance characteristics relative to a pre-trained diffusion model trained on ordinary PET data, as well as a superior lower bound for agreement with the 100\% count reconstructed image relative to existing methods.
\end{itemize}

\section{Background}

\newcommand{\imx}{\mathbf{x}}
\newcommand{\imq}{\mathbf{q}}
\newcommand{\imA}{\mathbf{A}}
\newcommand{\imy}{\mathbf{y}}
\newcommand{\imb}{\mathbf{b}}
\newcommand{\imB}{\mathbf{B}}
\newcommand{\imm}{\mathbf{m}}
\newcommand{\imz}{\mathbf{z}}
\newcommand{\score}{\nabla_{\imx} \log p_t(\imx_t) }
\newcommand{\scorecond}{\nabla_{\imx} \log p_t(\imx_t | \imm) }
\newcommand{\scorelike}{\nabla_{\imx} \log p_t(\imm | \imx_t) }
\newcommand{\scorenn}{ \mathbf{s}_{\mathbf{\theta}}(\imx_t, t)}
\newcommand{\stdnorm}{ \mathcal{N}(\mathbf{0}, \mathbf{I}) }
\newcommand{\tweedie}{\hat{\imx}_0(\imx_t)}
\newcommand{\xtone}[1]{\imx^{#1}_{t_{k+1}}}

\subsection{PET reconstruction}

Reconstructing an image from PET emission data is an inverse problem \cite{reader_deep_2021}. The mean $\imq$ of noisy measurements $\imm$ (e.g. a sinogram) may be modeled as 
\begin{equation}
    \imq = \imA\imx + \imb \;,
\end{equation}
where $\imx$ represents the true radiotracer distribution, $ \imA $ represents our system model and $ \imb $ models scatter and randoms components. The system model $\imA$ accounts for any positron range modeling, projection between image and sinogram space, as well as attenuation and normalization modeling.

PET emission data is generated as a set of random discrete emissions from radionuclides, and therefore follows a Poisson noise model. Maximum-likelihood expectation-maximization (MLEM) \cite{shepp_maximum_1982} is a convergent iterative algorithm that maximizes the Poisson log-likelihood (PLL) of emission data with respect to an image estimate, given by 
\begin{equation}
    L(\imx | \imm) = \sum_{i=1}^{k} m_i \log([\imA\imx + \imb]_i) - [\imA\imx + \imb]_i - \log(m_i!)\; .
\end{equation}
However, with low-count data and a high-dimensional image vector $\imx$, the maximum likelihood estimate overfits to measurement noise. It is standard to compensate for this reduction in signal by conditioning on an image-based prior, thereby regularizing the reconstruction, via algorithms such as maximum \textit{a posteriori}-expectation maximization (MAP-EM) \cite{levitan_maximum_1987}.

Such algorithms may be accelerated by partitioning sinograms into subsets and seeking the maxima of a set of corresponding sub-objectives, e.g. leading to Ordered-Subset Expectation Maximization (OSEM) \cite{hudson_accelerated_1994} and Block-Sequential Regularization Expectation Maximization (BSREM) \cite{de_pierro_fast_2001} for MLEM and MAP-EM respectively.

\subsection{MR-guided PET reconstruction}

MR-guided PET reconstruction can be achieved with a hand-crafted prior that promotes similarity between PET voxels whenever the corresponding voxels in the MR image are similar. The Bowsher method \cite{bowsher_utilizing_2004} is a well-established approach for achieving MR-based local smoothing. In this method, MAP-EM's prior penalty that encourages similarity between voxel intensities (e.g. quadratic, absolute) is weighted by the similarity of corresponding voxels in the MR image. 

In this work, we follow Schramm \textit{et al.} \cite{schramm_evaluation_2018}, using an asymmetrical relative difference prior (RDP) with Bowsher weighting using a patch-size of $3\times3\times3$ and a neighborhood size of 4 voxels.

\subsection{Score-based generative models (SGMs) / diffusion models}

Score-based generative models (SGMs) are a deep learning framework that may be employed to learn image probability distributions, given training images from the distribution \cite{ho_denoising_2020, sohl-dickstein_deep_2015, song_improved_2020}. Firstly, a diffusion process specifies how much Gaussian noise should be added to images at timestep $t$. Then, a time-conditional neural network $s_{\theta}$, called the score network, is trained to remove artificial Gaussian noise from training images, given the noisy image and associated timestep $t$. To sample new images from the learned distribution, an SGM samples a standard multivariate Gaussian to get $\imx_N$ and applies $s_{\theta}$ to get a sequence of images with ever-less artificial Gaussian noise $\imx_{N-1}, \imx_{N-2}, \dots, \imx_1, \imx_0$. This results in a high-quality image, sampled from the probability distribution of the training datasets.

In the rest of the work, we refer to diffusion models instead of SGMs for ease of readability.

\subsection{PET reconstruction with diffusion models --- PET-DDS}\label{sec:background-pet-dds}

We may train a diffusion model on 2D medical images and then utilize it as a learned prior for 3D medical image reconstruction \cite{chung_decomposed_2023, chung_solving_2023, webber_diffusion_2024}. Such approaches interleave the de-noising generative steps using $s_{\theta}$ with steps that encourage consistency with measured PET sinograms. In this work, we use the algorithm PET-DDS (Decomposed Diffusion Sampling) proposed by Singh \textit{et al.} \cite{singh_score-based_2024}.

In PET-DDS, data consistency steps constitute gradient descent on a sub-objective. This sub-objective is a weighted sum of the standard Poisson log-likelihood $L_j$ for subset $j$, an RDP penalty on the axial direction (introduced for consistency between neighboring 2D slices) and a third term to prevent the image iterate straying too far from the diffusion output. As done in previous work \cite{webber_likelihood_2025}, we introduce an additional hyperparameter $ \delta $ controlling the step size of gradient descent ($GD$) towards the reconstruction objective to ensure stable convergence towards the sub-objective.

\begin{algorithm}[t]
\caption{PET-DDS simplified pseudocode (one iteration)}\label{alg:pet-dds}
\begin{algorithmic}
    \Require Previous iteration $\xtone{ }$
    \Require Measurement-based normalization scale factor $c$
    \Require Anchoring strength $\lambda^{\text{DDS}}$
    \Require RDP strength $\lambda^{\text{RDP}}$
    \Require Gradient descent step size $\delta$
    \Require \# of gradient descent steps $p$
    \Require Number of subsets used $n_{\text{sub}}$
    \Require Current subset number $0 \le j < n_{\text{sub}} $
    \State $ \mathbf{x}^{(s)}_{t_k} \leftarrow s_{\theta}(\xtone{}, t_{k+1}) $ \Comment{Apply $s_{\theta}$ along 2D slices}
    \State $\xtone{0} \leftarrow $ Tweedie$(\mathbf{x}^{(s)}_{t_k}, t_k)$ \Comment{Estimate $\mathbb{E}(\imx_0 | \imx_{t_k})$}
    \State $\xtone{0} \leftarrow c^{-1} \cdot \xtone{0} $ \Comment{Scale according to measurements}
    \State $\Phi_j(\xtone{i}) \leftarrow L_j(\xtone{i}) + \frac{1}{n_{\text{sub}}} \bigl( \lambda^{\text{RDP}} RDP_z(\xtone{i}) $
    \State $\qquad -\lambda^{\text{DDS}} ||\xtone{i}  - \xtone{0}||^2_2 \bigr)$ \Comment{Define sub-objective}
    \For{\textit{i = 1,..., p}} $ \xtone{i} \leftarrow GD(\delta \cdot \Phi_{j+i}(\xtone{i-1}))$
    \EndFor
    \State $ \xtone{p} \leftarrow c \cdot \xtone{p}$ \Comment{Normalize image to diffusion scale}
    \State $\imx_{t_k} \leftarrow $ Re-noise$(\mathbf{x}^{(s)}_{t_k}, \xtone{p}, \mathcal{N}(0, \mathrm{I})) $ \Comment{Re-noising step}
\end{algorithmic}
\end{algorithm}

The full algorithm also introduces measurement-based normalization. The diffusion model is trained on normalized images, and so normalization steps are also introduced to transform image intensities between the domain of the diffusion model and the domain of the Poisson log-likelihood function. Algorithm \ref{alg:pet-dds} gives a simplified pseudocode. For the full mathematical details, see Singh \textit{et al.} \cite{singh_score-based_2024}.

The authors of PET-DDS also propose an extension of their method to MR-guided PET reconstruction, which we call here PET-DDS + MR. This algorithm trains a conditional diffusion model on paired PET-MR images using classifier-free guidance \cite{ho_classifier-free_2022}, before then including the MR image as an input to $s_{\theta}$ during reconstruction.

\subsection{Deep-learned image registration}

Image registration is the problem of finding an optimal spatial transformation from a ``moving'' image $\mathbf{x}$ to a ``fixed'' image $\mathbf{f}$.
For affine registration, we seek the affine map $A$ that minimizes the loss $\mathcal{L}(\mathbf{f}, A\mathbf{x})$.
More generally, we seek a nonlinear deformation field $\phi$ that minimizes $\mathcal{L}(\mathbf{f}, \mathbf{x} \circ \phi)$.
VoxelMorph \cite{balakrishnan_voxelmorph_2019} is a deep learning framework that parameterizes the deformation field by its inputs $ \mathbf{f} $ and $ \mathbf{x} $, such that we seek to learn $\phi(\cdot, \cdot)$ given a training dataset. The deformation function $\phi(\cdot, \cdot)$ is based on a U-Net architecture, and outputs the deformation field between inputs $ \mathbf{f} $ and $ \mathbf{x} $ (i.e. to transform a vector $\mathbf{z} $ from the space defined by $\mathbf{x}$ to the space defined by $\mathbf{f}$, we compute $\phi(\mathbf{f}, \mathbf{x})(\mathbf{z})$). Its training loss sums a voxel-wise registration error $ || \mathbf{f}- \mathbf{x} \circ \phi(\mathbf{f},\mathbf{x}) ||_2^2$ and a smoothness penalty on $\phi(\mathbf{f},\mathbf{x})$.

\section{Related work}\label{sec:related_work}

Historically, approaches for MR-guided PET reconstruction have posed and solved a regularized reconstruction problem based on hand-crafted priors that encourage some similarity between the reconstructed image and the MR guide image \cite{bai_mr_2013, bowsher_utilizing_2004, bland_mr-guided_2018}. Such approaches make the implicit assumption that similar structures in the MR image should correspond to similarities within the PET image, which can lead to artifacts at high regularization strengths.

Deep-learning approaches to MR-guided reconstruction have been proposed to learn to make the best use of MR images to improve the quality of reconstructed PET images \cite{mehranian_model-based_2020, xie_anatomically_2021, mehranian_pet_2017}. However, supervised deep learning approaches can suffer from over-fitting and the corresponding lack of generalization ability unless many datasets are available \cite{costa-luis_micro-networks_2021}.

A different approach was taken by Gong \textit{et al.}, who proposed using the unsupervised deep image prior method for PET reconstruction. MR conditioning was achieved by using the MR guide image as the starting point for the deep image prior optimization \cite{gong_pet_2019}.

As mentioned in the introduction, Singh \textit{et al.} proposed utilizing only images rather than full sinogram data for training a flexible reconstruction algorithm for PET reconstruction based on diffusion models \cite{singh_score-based_2024}. This showed improved lesion detectability compared to the deep image prior for out-of-distribution lesions. The authors also proposed an extension to MR-guided reconstruction by training the diffusion model on paired PET-MR images. Compared to the PET-only algorithm, they showed improvements in reconstruction accuracy metrics such as peak signal-to-noise ratio (PSNR) and structural similarity index measure (SSIM) at the cost of worse contrast recovery coefficient (CRC).

Registration of PET images to a specific subject \textit{post-reconstruction} has been previously investigated, for example Burgos \textit{et al.} showed benefits for detecting metabolic abnormalities from FDG PET \cite{burgos_subject-specific_2015}.

The work presented in this article builds on work by Singh \textit{et al.} \cite{singh_score-based_2024}, and also substantially extends earlier work proposed at the IEEE Medical Imaging Conference 2024 \cite{webber_multi-subject_2024}.

\section{Methods}

\subsection{Approach}

\begin{figure*}
    \centering
    \includegraphics[width=1\linewidth]{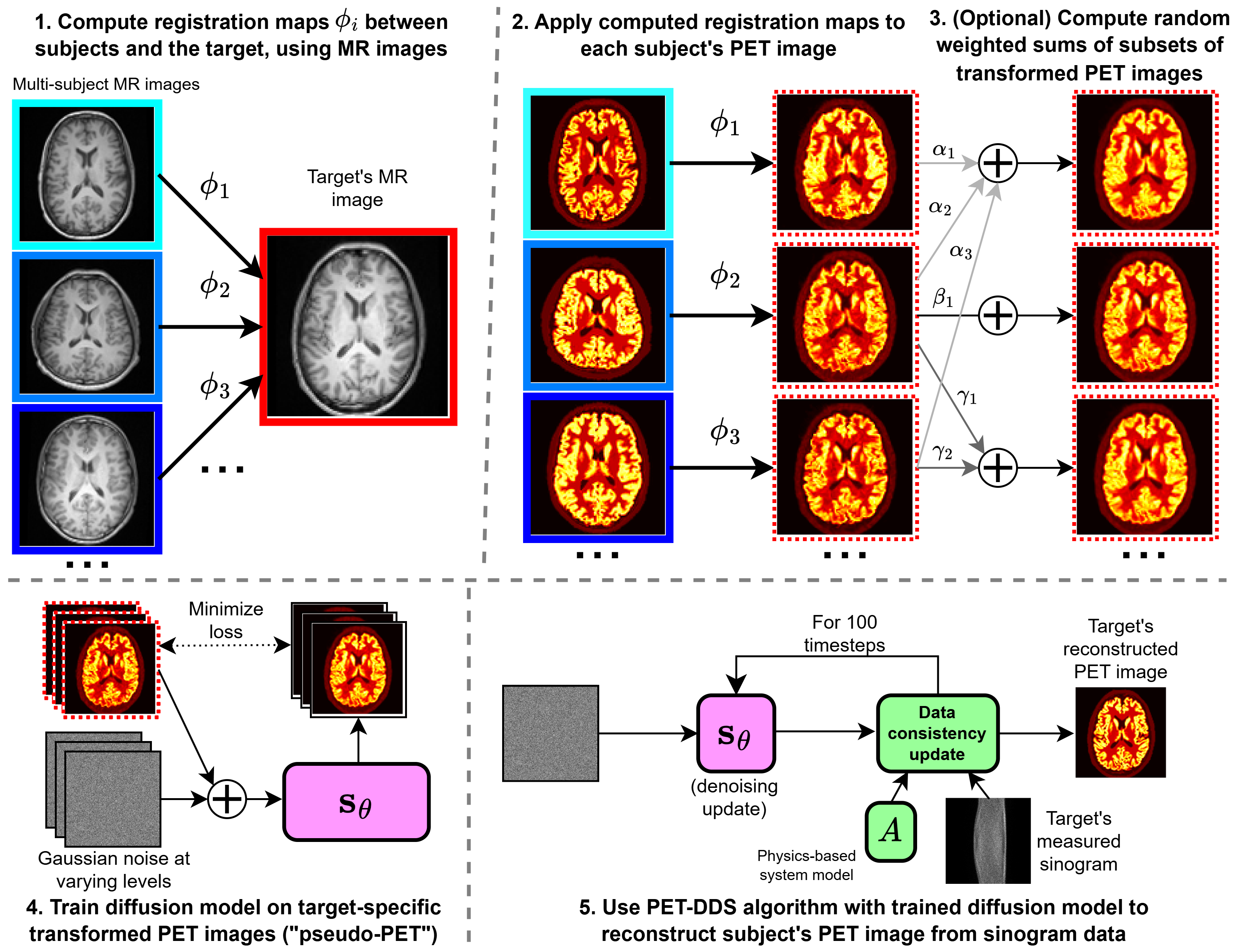}
    \caption{Outline of our reconstruction methodology to reconstruct the target's PET image: generating pseudo-PET images in our single-subject target space (steps 1, 2 \& 3); incorporating target-specific pseudo-PET images as a prior for diffusion-model-based reconstruction (step 4); and, reconstructing from target PET data using the personalized diffusion model and the PET-DDS algorithm (step 5). In steps 1-3, images with red borders are in the single-subject target space, images with blue borders are in their original untransformed space, while images with dashed borders have undergone non-linear registration.}
    \label{fig:methodology}
\end{figure*}

We consider the case where we have a dataset of previously acquired paired PET-MR images $ \left( \imx_i^{\text{PET}}, \imx_i^{\text{MR}} \right)_{i=1}^{N}$. Our goal is to reconstruct the target's PET image, from their measured sinogram $\imm$ and MR image $\imx^{\text{MR}}_{\text{target}}$. To this end, we seek to generate pseudo-PET images that are in the anatomical space defined by our target subject, and thereby personalize an SGM for reconstruction. Hypothetically, this personalized diffusion model has an easier learning task than a non-personalized diffusion model, as it only needs to learn PET distributions in the target anatomy (instead of generalizing to unseen brain anatomies). The model is also trained on less-noisy pseudo-PET images, that have anatomical features of relevance, so the prior information being incorporated is of greater relevance. See Figure \ref{fig:methodology} for a graphical overview of our approach. 

\subsubsection{Generation of pseudo-PET (see Fig. \ref{fig:methodology}.1-\ref{fig:methodology}.3)}\label{sec:methods_approach_ppgen}

We begin by resampling and cropping MR images to the PET voxel size and region of interest in the scanner.

For each image pair $\imx_i$ in our dataset, we then perform affine registration of $\imx_i^{\text{MR}}$ to the same affine space (defined relative to a fixed MR image $ \imx^{\text{MR}}_{\text{affine}} $, which may be the subject's MR image but does not have to be), using NiftyReg \cite{modat_global_2014}. This yields an affine matrix $\imB_i$ for each image.

We then train a VoxelMorph instance $ \phi $ to minimize the loss $\ell (\imB_{\text{target}}\imx^{\text{MR}}_{\text{target}}, \cdot) $ on the affine-registered datasets $ \left( \imB_i \imx_i^{\text{MR}} \right)_{i=1}^{N} $.

Using the trained VoxelMorph instance $\phi$, we compute the VoxelMorph registration map for image $i$ as $\phi_i := \phi(\imB_{\text{target}}\imx^{\text{MR}}_{\text{target}}, \imB_i\imx_i^{\text{MR}})$, and use it to transform PET image $\imx_i^{\text{PET}}$ to the target reference space by computing $ \imB_{\text{target}}^{-1}\phi_i(\imB_i\imx_i^{\text{PET}})$. This image is therefore a mapping of the radiotracer uptake (as measured by PET) in one patient's brain to the anatomy of a different patient's brain.   

As all pseudo-PET images here belong in the same reference space, we may choose to sum multiple transformed PET images together to increase the diversity and statistical quality of our dataset. For this purpose, we (1) choose the number of images to sum $N$ such that $\mathbb{P}(N=n) \propto \frac{1}{n} $, (2) choose $N$ images randomly without replacement, (3) sample weights for each image as scalars $w_i \sim U[0,1]$, (4) scale the weights $(w_i)$ such that they sum to 1, and (5) sum the weighted images. This method of randomly combining the images is heuristically proposed to balance the potential benefit of learning from less-noisy images (i.e. multiple images summed together) and the potential risk that all images appear too similar (if too many images are summed each time and no images in the resulting dataset have fine-grained features).

\subsubsection{Training diffusion model with pseudo-PET (see Fig. \ref{fig:methodology}.4)}

We followed the approach given by Singh \textit{et al.} for training a diffusion model to use for 3D PET-DDS reconstruction \cite{singh_score-based_2024}. Training images were first normalized with Singh \textit{et al.}'s measurement-intensity normalization procedure, and empty transverse slices were removed. During training, the global magnitude of each slice was varied for each epoch by dividing by a value sampled uniformly between 0.5 and 1.5.

Because all slices are co-registered to the target reconstruction space in advance, we use slice conditioning (with the slice number expressed as a value between 0 and 1) as an additional input to our diffusion model, which is embedded using the same neural architecture as the timestep embedding.

All diffusion models were trained with lesion-free data, to simulate the difficult case of addressing out-of-distribution PET-MR mismatches at test time. 

\subsubsection{Reconstruction with pre-trained diffusion model (see Fig. \ref{fig:methodology}.5)}

The PET-DDS algorithm was used, with the modification that the diffusion model was also slice-conditioned (applied as an additional embedding alongside the usual timestep embedding).

\subsection{Data}

\subsubsection{Simulated \textnormal{\FDG} data}

To evaluate our method's reconstruction accuracy, we simulated ground truth \FDG PET scans (and corresponding attenuation maps) from 39 real T1 MR images (following \cite{mehranian_model-based_2020}) using representative activity values for segmented tissues.
Poisson noisy sinograms at 100\% dose were simulated at $6 \times 10^8 $ counts per 3D volume (a clinical equivalent dose).

Where algorithms require image-based training data, OSEM reconstructions from these 100\% dose sinograms were used for training. For evaluating reconstruction methods, Poisson noisy sinograms were used with 2.5\% of clinical counts.

Attenuation was modeled in the simulated data and included in the reconstructions, along with a constant background (to represent scatter and randoms) equal to 30\% of the count level for the prompts (at a given dose).

For reconstruction (and not training), hot lesions were simulated as uniform spheres with radii between 2 mm and 10 mm with an activity level of $1.5 \times$ gray matter. Lesions were simulated at MR resolution and down-sampled to PET resolution \cite{mehranian_model-based_2020}. Sinograms were then computed as before, except now for a ground truth containing lesions.

\subsubsection{Real \FDG data}

In addition to simulation studies, we tested our approach with real \FDG brain datasets. 36 static datasets were used in this work (see \cite{mehranian_model-based_2020} for a full description).

The data had been previously acquired from 30-minute to one-hour scans with the Siemens Biograph mMR, with approximately 200 MBq administered, with total counts in the range $1.2 \times 10^8 $ to $ 9.1 \times 10^8$. At full count, high-quality images (voxel size 2.08626 mm $\times$ 2.08626 mm  $\times$ 2.03125 mm) were reconstructed with the scanner defaults (OSEM with 21 subsets, 2 iterations, and no positron range modeling; 3D image size $344 \times 344 \times 127$). The images were then cropped to $128 \times 128 \times 120$ voxels, to remove empty space.

For each dataset, Siemens' scanner-specific algorithms were used to produce normalization sinograms, compute attenuation maps from previously acquired paired CT scans, and approximate the distribution of scatter and randoms events.

To simulate low-count data, counts were sampled without replacement from the prompt sinogram such that in expectation each prompt sinogram bin had a mean count rate $40\times$ lower than the full-count sinogram (i.e. 2.5\% of full-count). Siemens' scanner-specific algorithms were then used to compute contamination sinograms.

\subsection{Comparison reconstruction algorithms}

\subsubsection{OSEM}

OSEM is the standard reconstruction algorithm used clinically. We fixed the number of subsets at 21 for this work, with the number of iterations varied as a hyperparameter.

\subsubsection{Bowsher}

Block-sequential regularized expectation maximization (BSREM) is an iterative algorithm for solving a regularized tomographic expectation maximization problem by splitting the tomographic views into subsets for efficient evaluation. We follow Schramm \textit{et al.}'s \cite{schramm_evaluation_2018} recommendation to use the asymmetric Bowsher prior with RDP penalty as our regularization function providing anatomical guidance. The algorithm we use therefore maximizes the following objective:

\begin{equation}
     L(\imx | \imm ) + \beta R_{aBOW}(\imx) \; ,
\end{equation}
 where $R_{aBOW}$ is the asymmetric Bowsher prior.

The key hyperparameters we vary are $\beta$, controlling the strength of regularization on MR data, and the number of iterations taken. Other hyperparameters for $R_{aBOW}$ were taken to match those found by Schramm \textit{et al.} \cite{schramm_evaluation_2018}.

In the case of simulated data, excluding inserted lesions, the PET ground truth was analytically derived from MR data. As a result, the Bowsher prior is an unrealistically good prior for this specific dataset and may be thought of as an upper bound on what is possible with a perfect handcrafted prior.

\subsubsection{Diffusion model methods}
As discussed in Sections \ref{sec:background-pet-dds} and \ref{sec:related_work}, Singh \textit{et al.} previously showed state-of-the-art results for unsupervised PET reconstruction (i.e. with no sinogram data available for training) with their algorithm PET-DDS \cite{singh_score-based_2024}.

We denote their original algorithm by ``PET-DDS", and the version incorporating MR information with classifier-free guidance as ``PET-DDS + MR", where $\lambda^{MR}$ denotes the guidance strength. For PET-DDS + MR, the probability of unconditional training used was 0.1.

We refer to PET-DDS trained with multi-subject PET data that has been registered to the target's reference space as ``MR-reg + PET-DDS", while we refer to PET-DDS trained with data that has been registered and summed together (see Section \ref{sec:methods_approach_ppgen}) as ``MR-reg \& sum + PET-DDS".

For the diffusion-model-based approaches, we fix the gradient descent step size as $ \delta = 0.2 $ and vary the strength of regularization imposed by the diffusion model by varying the hyperparameter $\lambda^{\text{DDS}}$. For non-registered training data, random affine augmentation (uniformly sampling up to a $5\%$ scaling, up to a 5 voxel translation, and up to a $5^\circ$ rotation) was applied to images during training, to enhance generalization to unseen datasets.

We perform 100 diffusion steps per reconstruction, 10 data consistency steps per diffusion step, and fix the hyperparameter $\lambda^{\text{RDP}} = 5 \times 10^{-4}$. Similar to OSEM and Bowsher, we use 21 subsets for data consistency updates.

\subsection{Further experimental details}
\subsubsection{Diffusion model architecture}
Following Singh \textit{et al.} \cite{singh_score-based_2024}, diffusion models were trained with the architecture proposed by Dhariwal \& Nichol \cite{dhariwal_diffusion_2021}, which is based on a U-Net architecture \cite{ronneberger_u-net_2015} with global attention layers at the lowest spatial sampling resolution ($8 \times 8$).

\subsubsection{Training procedures}
For both simulated and real data, diffusion models were trained on the denoising score-matching loss \cite{vincent_connection_2011}, with batch size 16 and optimized with Adam with learning rate $1\times 10^{-4}$. Three validation datasets were used to select the parameters that best minimize the validation loss. For MR-reg \& sum + PET-DDS, for each training epoch, a new set of pseudo-PET images was generated online (of equal size to the dataset of PET images used by other methods). All models were implemented in PyTorch, and all experiments were conducted on an NVIDIA GeForce RTX 3090 with 24 GB GPU memory.

\subsubsection{Registration function training}
The VoxelMorph instances trained for both simulated and real data were trained for 1500 epochs, with the normalized cross-correlation loss with no regularization. Losses were evaluated on registration to the target dataset only. Over-fitting is not a concern for this task, as the test data is available during training.

\subsubsection{PET forward operator}

Each reconstruction method made use of the same 3D ParallelProj projector
\cite{schramm_parallelprojopen-source_2024}, with span 11 compression and without positron range modeling. Further details of how the scanner geometry was modeled are given in Webber \textit{et al.} \cite{webber_likelihood_2025}.

\subsubsection{Bias-variance trade-off calculation}

 To calculate the 2D bias-variance assessment in Section \ref{sec:results_simrecon}, for each of 10 random seeds, noisy sinogram data was generated according to the Poisson noise model. Then, reconstructions were performed for each independent noisy realization, with bias and standard deviation calculated according to Reader \& Ellis \cite{reader_bootstrap-optimised_2020}. The OSEM reconstruction from full-count data was taken as the ground truth for this analysis.

\section{Results}

\subsection{Example pseudo-PET images}

\begin{figure}
    \centering
    \includegraphics[width=1\linewidth]{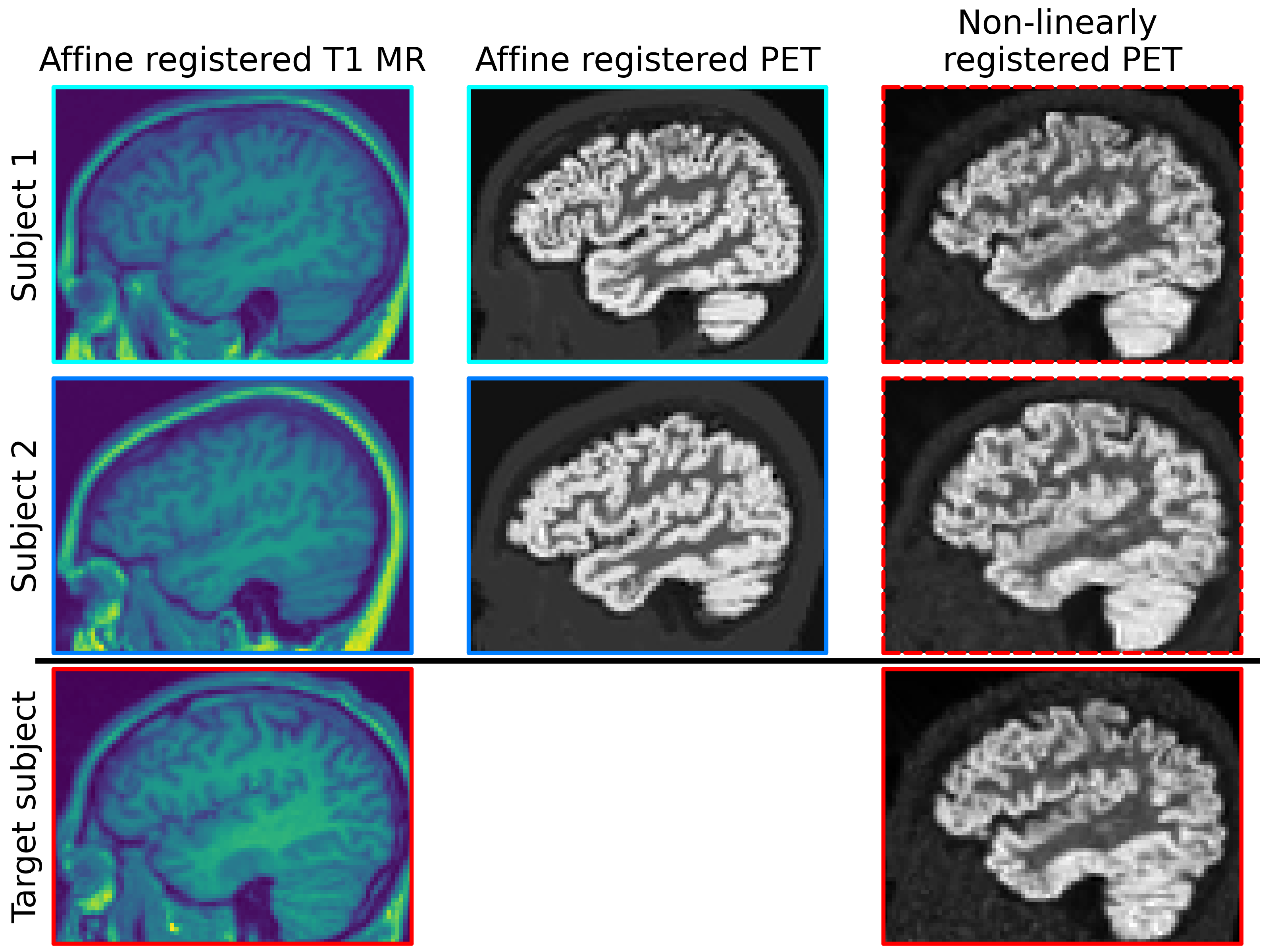}
    \caption{Simulated data: example sagittal slices from 3D pseudo-PET images that have been transformed via deep-learned MR-to-MR registration maps, with the clinical PET and MR images for comparison. The PET and MR images in columns 1 and 2 are shown after the initial affine registration stage. Red = image in target space; blues = rigid transformations of images from their original spaces; dashed line = non-linearly registered images.}
    \label{fig:sim_pseudo_pet}
\end{figure}

\begin{figure}
    \centering
    \includegraphics[width=1\linewidth]{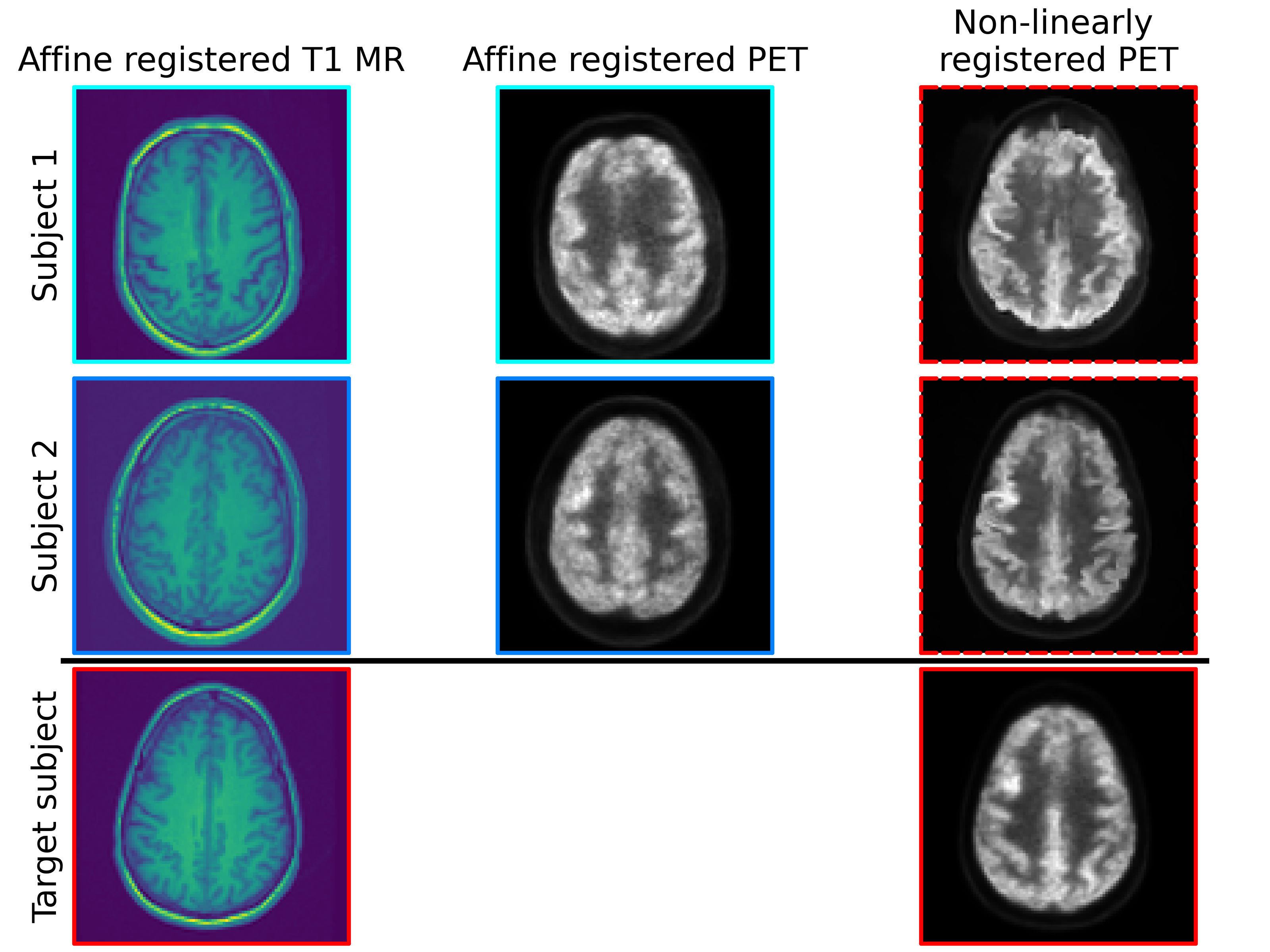}
    \caption{Real \FDG data: example transverse slices from 3D pseudo-PET images that have been transformed via deep-learned MR-to-MR registration maps, with the clinical PET and MR images for comparison. The PET and MR images in columns 1 and 2 are shown after the initial affine registration stage. Red = image in target space; blues = rigid transformations of images from their original spaces; dashed line = non-linearly registered images.}
    \label{fig:real_pseudo_pet}
\end{figure}

In Figure \ref{fig:sim_pseudo_pet}, we show example pseudo-PET images for the simulated case. In Figure \ref{fig:real_pseudo_pet}, we show example pseudo-PET images for real \FDG data.

In both cases, we observe that pseudo-PET images retain some of the anatomical information from the target MR image. In particular, in Figure \ref{fig:real_pseudo_pet} we can see the distribution of white and gray matter in the transformed PET closely resembles the target reconstruction. However, there are still differences between the two transformed PET images, so it is plausible that using a set of these images as training data will allow the diffusion model to learn a wide enough image probability distribution to perform well at reconstruction tasks.

\subsection{Simulated data reconstruction}\label{sec:results_simrecon}

\begin{figure*}
    \centering
    \includegraphics[width=1\linewidth]{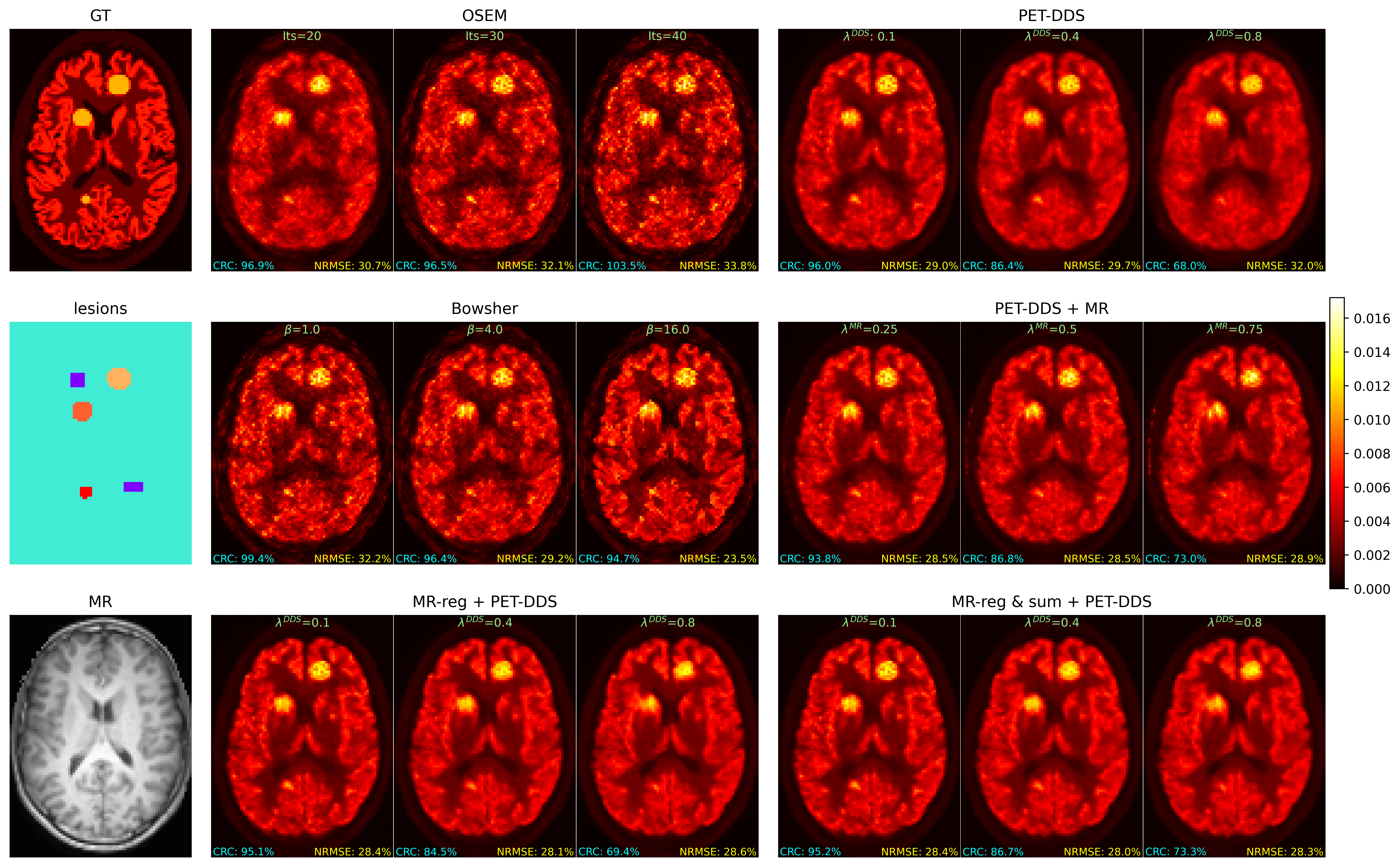}
    \caption{Reconstructed transverse image slices for each of six PET reconstruction algorithms (each shown with 3 different hyperparameter choices corresponding to different regularization strengths) from a 3D dataset with out-of-distribution lesions. The contrast recovery coefficient (CRC) was calculated using the purple rectangular regions as background regions (corresponding to uniform white matter in the ground truth) and averaging the CRC for each of the three shown hot lesions. NRMSE was evaluated on pixels that were non-zero in the ground truth. Hyperparameters for each method were selected to show a representative cross-section of performance at different regularization strengths. MR-reg \& sum + PET-DDS performed similarly to MR-reg + PET-DDS, and so is omitted for real data studies for brevity.}
    \label{fig:sim_images}
\end{figure*}

We present results evaluating the trade-off between reconstruction performance on areas of PET-MR mismatch (i.e. out-of-distribution PET-specific lesions) and areas of agreement (i.e. within-distribution brain structures), for reconstructions from 2.5\% of full-count data. Figure \ref{fig:sim_images} shows this trade-off qualitatively for a range of hyperparameter choices for each considered algorithm. From this figure, we can see that our proposed approaches (bottom row) can visually recover lesions better than PET-DDS and PET-DDS + MR while retaining the improved reconstruction accuracy associated with MR guidance. As expected, OSEM reconstructions are quantitatively accurate (i.e. good CRC scores on large areas due to low bias), but failure to recover fine details in the presence of noise makes lesion detection qualitatively harder with more iterations. Bowsher reconstructions perform very well, in part due to the strong link between MR and PET for this simulated data. However, lesions appear blended with MR-specific anatomy at higher penalty strengths (this effect is also observed for PET-DDS + MR), with this effect particularly pronounced for the leftmost lesion that straddles both the striatum and white matter.

\begin{figure}
    \centering
    \includegraphics[width=1\linewidth]{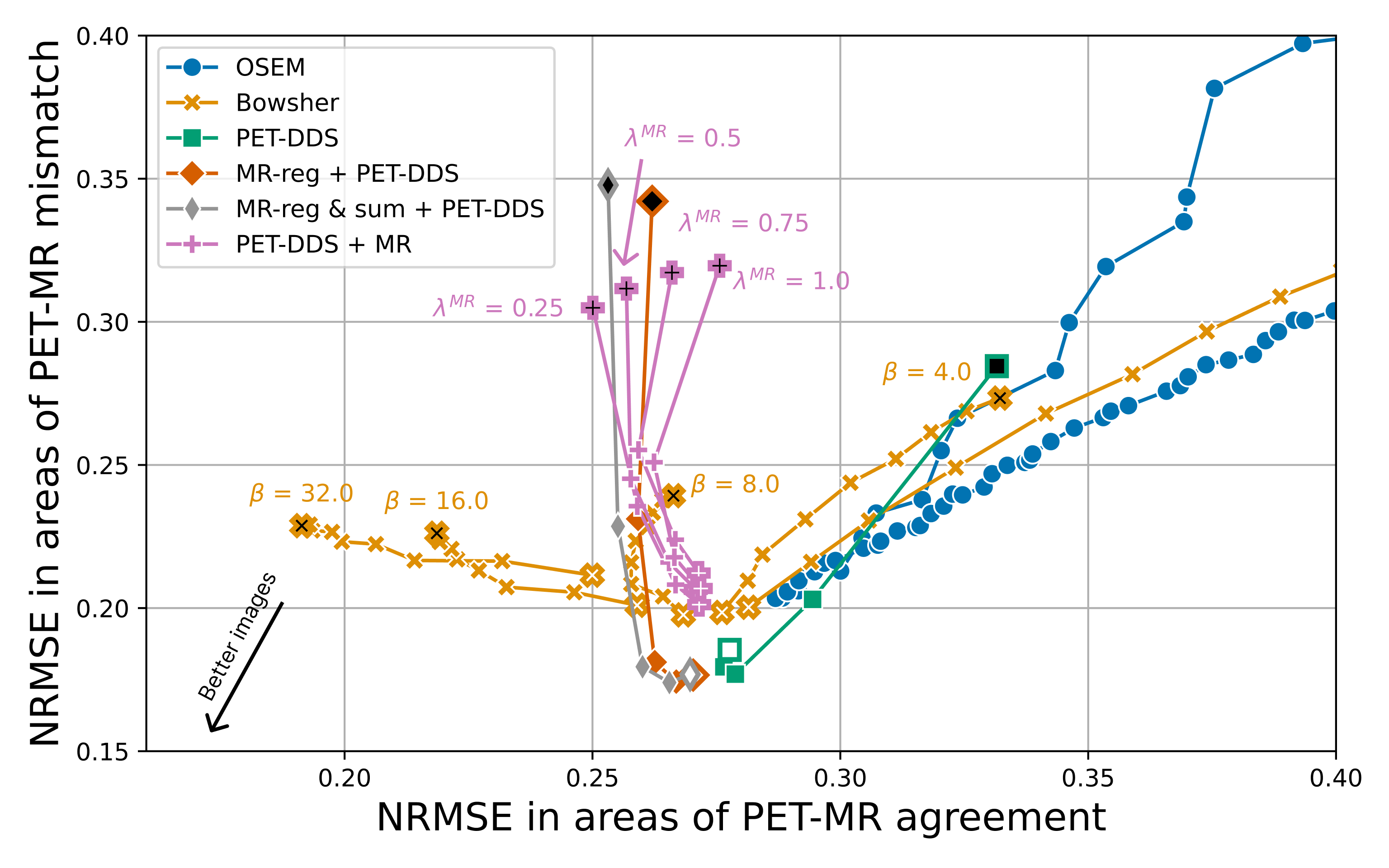}
    \caption{Error trade-off between areas of PET-MR agreement and areas of PET-MR mismatch (i.e. PET-only lesions) for different reconstruction algorithms. NRMSE in areas of PET-MR mismatch was calculated in lesion pixels only, while NRMSE in areas of PET-MR agreement was calculated across a set of $10 \times 10 \times 10$ patches of representative non-lesion brain (one per lesion). For OSEM and Bowsher, white markers correspond to the lowest number of iterations (5) and black markers correspond to the maximum number of iterations. For PET-DDS-based methods, white markers correspond to the smallest value of $\lambda^{\text{DDS}}$, with black markers corresponding to the greatest value of $\lambda^{\text{DDS}}$, reflecting increased guidance strength from the pre-trained diffusion model.}
    \label{fig:sim_nrmse_mismatches}
\end{figure}

\begin{figure}
    \centering
    \includegraphics[width=1\linewidth]{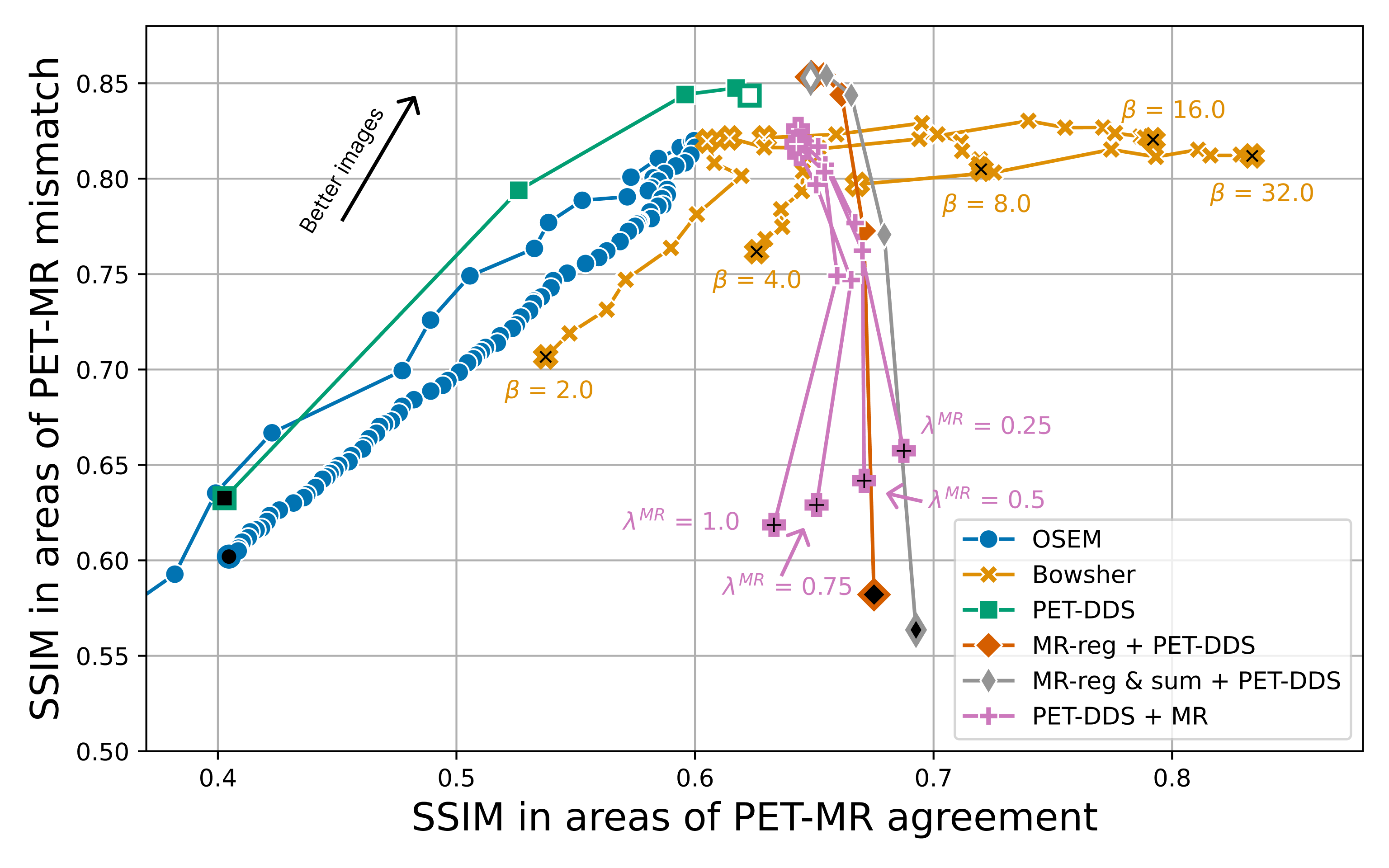}
    \caption{Structural similarity trade-off between areas of PET-MR agreement and areas of PET-MR mismatch (i.e. PET-only lesions) for different reconstruction algorithms. SSIM in areas of PET-MR mismatch was calculated as the average SSIM in $10 \times 10 \times 10$ patches containing lesions, while SSIM in areas of PET-MR agreement was calculated as the average SSIM in a set of $10 \times 10 \times 10$ patches of representative non-lesion brain voxels (one per lesion). See Figure \ref{fig:sim_nrmse_mismatches} for marker meanings.}
    \label{fig:sim_ssim_mismatches}
\end{figure}

For more detailed quantitative analysis, in Figures \ref{fig:sim_nrmse_mismatches} and \ref{fig:sim_ssim_mismatches} we evaluate the NRMSE and SSIM of areas containing PET-MR mismatches against those that only contain PET-MR agreement. Our findings agree with Figure \ref{fig:sim_images}, and in particular, show higher peak reconstruction accuracy in areas of PET-MR mismatch for our proposed methods compared to all other methods. Similarly, at a given level of reconstruction accuracy in areas of PET-MR mismatch (i.e. a proxy for lesion detectability), our approach outperforms PET-DDS and PET-DDS + MR on reconstruction accuracy in areas of PET-MR agreement.

\begin{figure}
    \centering
    \includegraphics[width=1\linewidth]{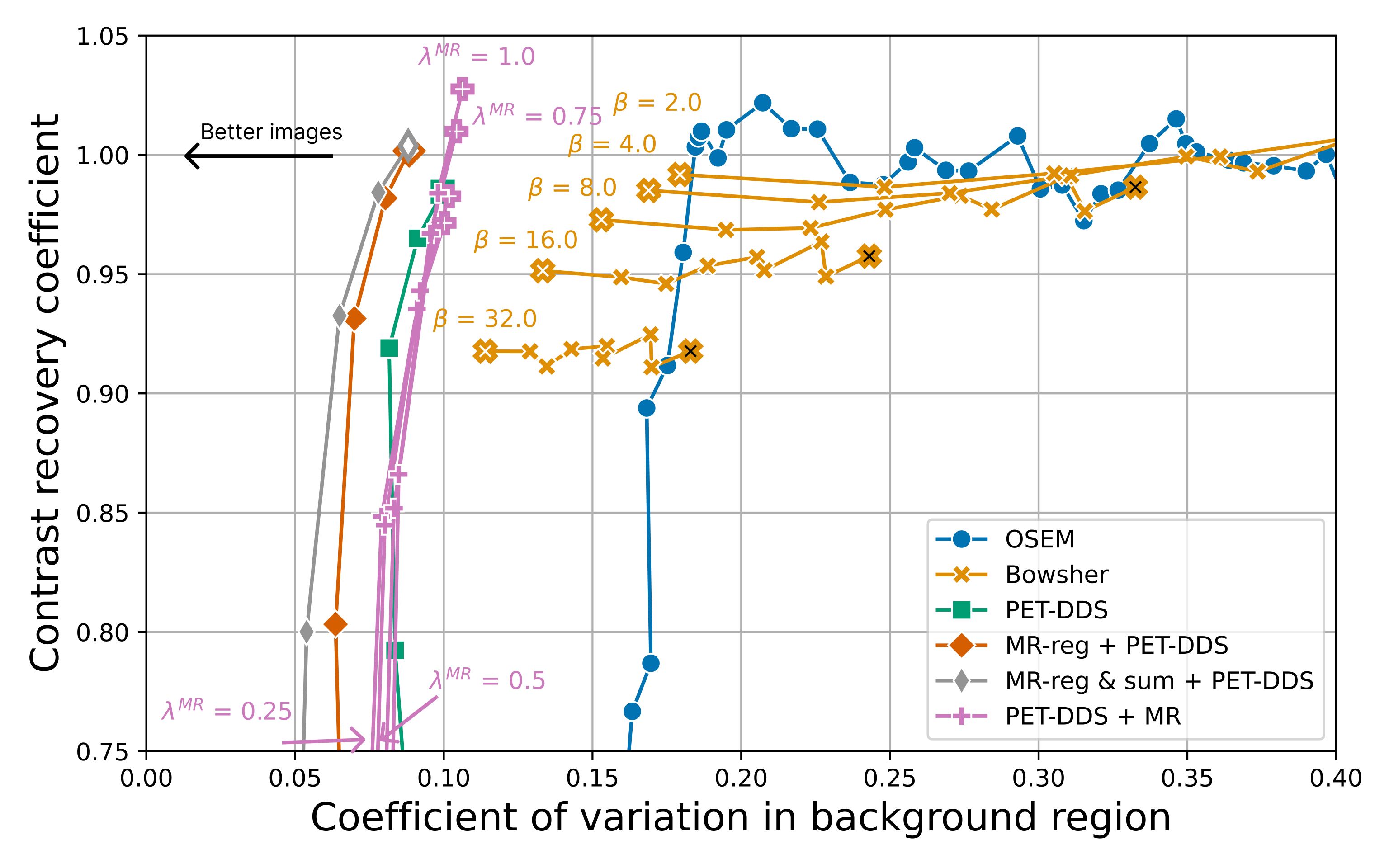}
    \caption{Trade-off between lesion contrast recovery coefficient (CRC) and noise in background area for different reconstruction algorithms. CRC was calculated as the average CRC for each lesion, measured against a background region of white matter (uniform in ground truth), with the coefficient of variation measured in the same background region. See Figure \ref{fig:sim_nrmse_mismatches} for marker meanings.}
    \label{fig:sim_crc_mismatches}
\end{figure}

We also compared the lesion detectability (measuring the CRC across all lesions) against the background noise in the reconstruction for each method in Figure \ref{fig:sim_crc_mismatches}. This figure shows that our proposed approach yields lower background noise for a given CRC value than other approaches.

As expected, reconstruction from MR-reg \& sum + PET-DDS is marginally smoother than MR-reg + PET-DDS, due to the smoothing effect of summing together training data. However, overall MR-reg \& sum + PET-DDS performed similarly to MR-reg + PET-DDS, and so is omitted from our real data studies for brevity.

\subsection{Algorithm variants}\label{sec:results_variants}
\subsubsection{PET-PET registration for pseudo-PET images}

\begin{figure*}[htbp]
    \centering
    \begin{subfigure}[b]{0.53\linewidth}
        \includegraphics[width=1\linewidth]{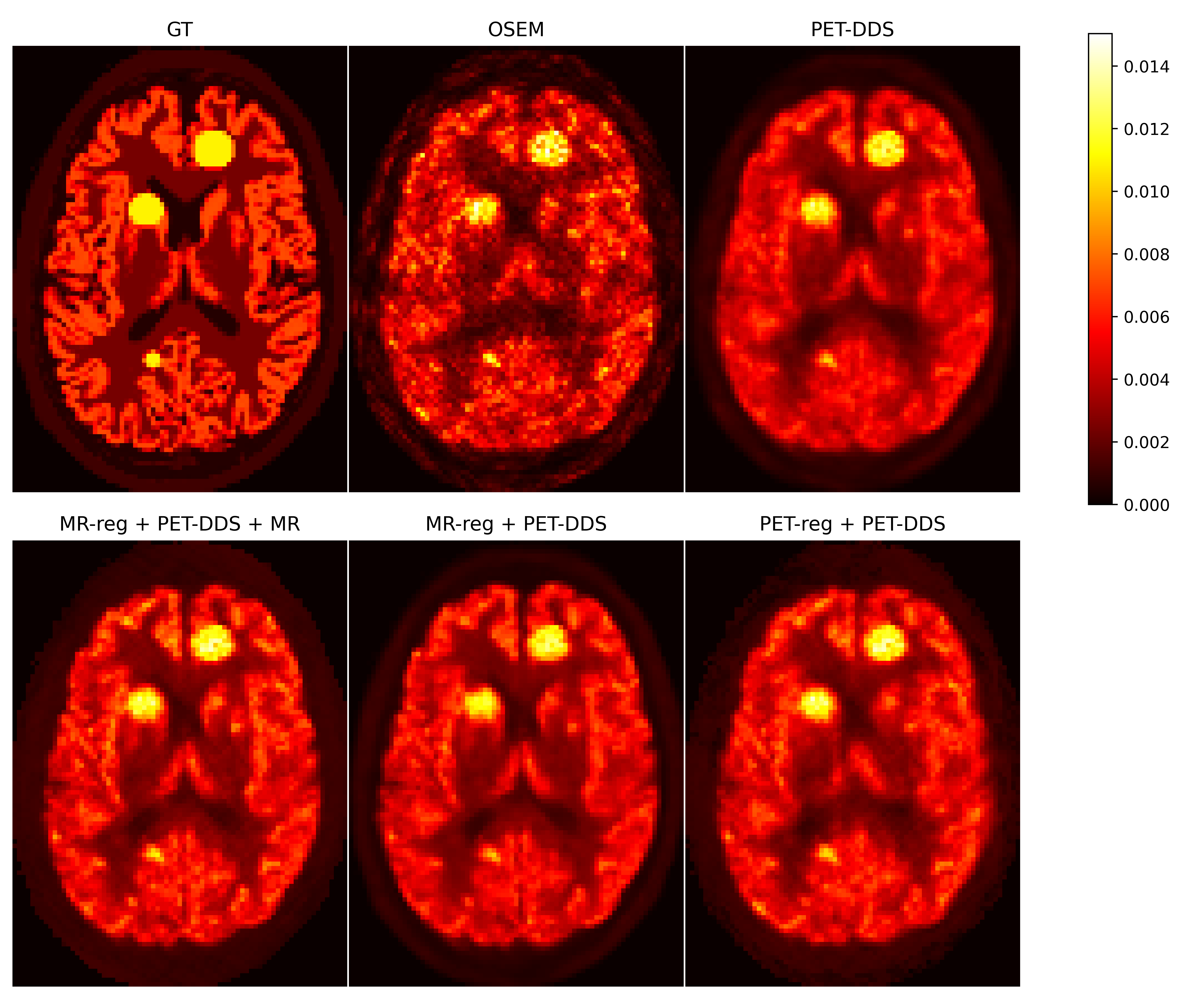}
        \caption{Example reconstruction of the algorithm variants PET-reg + PET-DDS and MR-reg + PET-DDS + MR. All SGM-based methods used the same hyperparameter $\lambda^{DDS} = 0.4$.}
        \label{fig:pet_pet_images}
    \end{subfigure}
    \hspace{0.02\textwidth}
    \begin{subfigure}[b]{0.42\linewidth}
        \includegraphics[width=\linewidth]{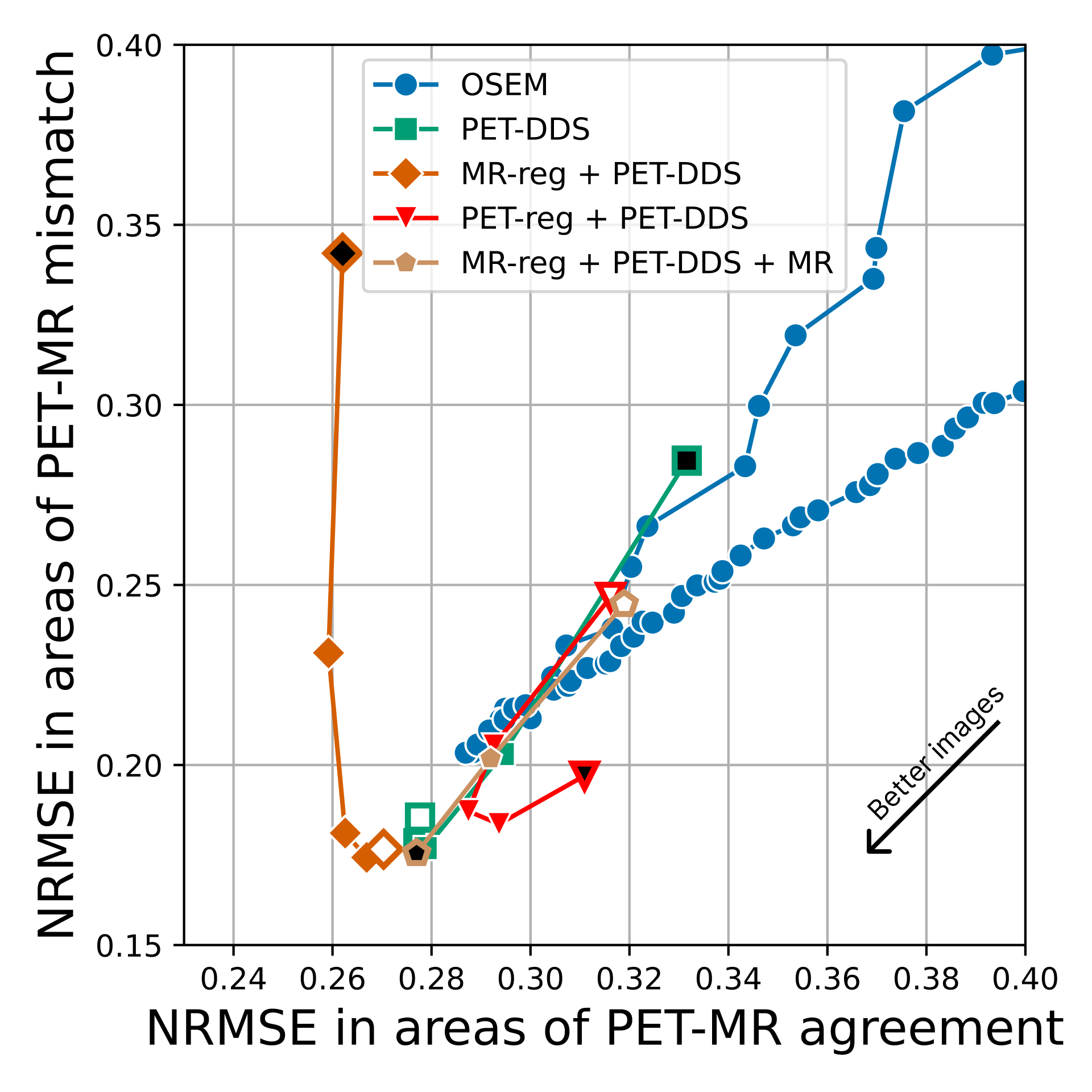}
        \caption{Error trade-off between areas of PET-MR agreement and areas of PET-MR mismatch (i.e. PET-only lesions) for selected algorithms. (See Figure \ref{fig:sim_nrmse_mismatches} for further details.)}
        \label{fig:pet_pet_stats}
    \end{subfigure}
    \caption{Results (on simulated data at 2.5\% count) for variants of the proposed method. PET-reg + PET-DDS is a variant of the proposed method that uses PET-based registration to generate pseudo-PET images. MR-reg + PET-DDS + MR is a variant that uses MR-registered pseudo-PET images, and additionally provides the subject's MR image as a side-channel to the SGM (during training and reconstruction).}
    \label{fig:pet-pet}
\end{figure*}

We investigated using PET-PET registration instead of MR-MR registration for producing pseudo-PET images. In Figure \ref{fig:pet_pet_images}, we show example quantitative and qualitative results from the method `PET-reg + PET-DDS` (PET-DDS trained with PET-registered pseudo-PET images). This figure shows that reconstructions from an SGM trained on PET-registered images have fewer anatomical details present than their MR-registered equivalent. Compared to standard PET-DDS, background noise appears less well suppressed. While some structures appear clearer than in PET-DDS (e.g. the cortical folds at the top of the image), others appear more hallucinatory (e.g. the thalamus, appearing as a central inverted v-shape). These observations are bourne out by the quantitative performance of the method, summarized in Figure \ref{fig:pet_pet_images}.

\subsubsection{Including MR information as an additional channel for MR-reg + PET-DDS}

We also investigated supplying the subject's MR image to the SGM during training and reconstruction. In our experiments, the MR image was supplied as a second channel to the PET image. Figure \ref{fig:pet-pet} shows the qualitative and quantitative results from experiments with this variant, demonstrating that there seems to be no additional benefit to including the MR-information in this secondary way.

\subsection{Real data reconstruction}

\begin{figure}
    \centering
    \includegraphics[width=\linewidth]{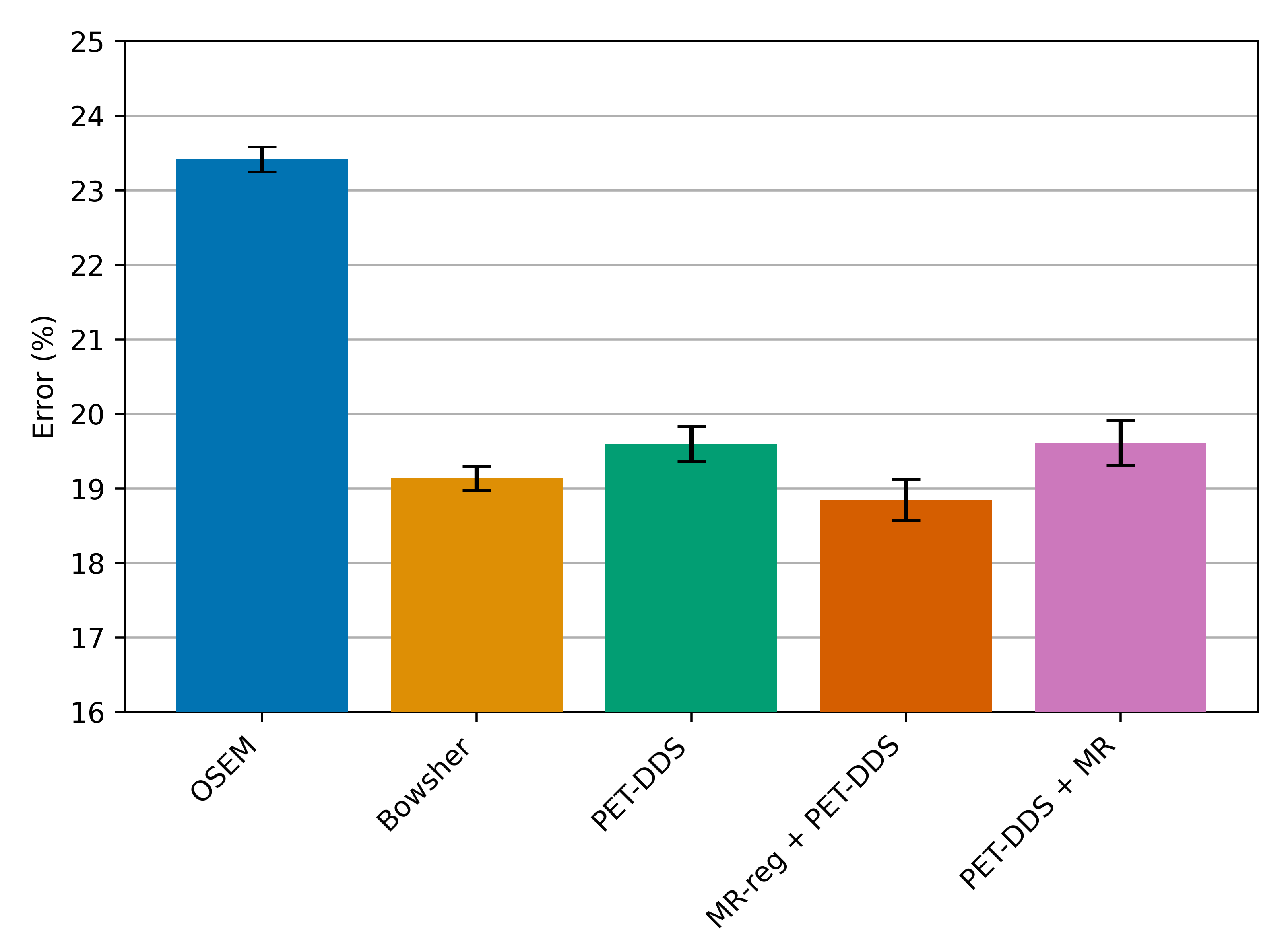}
    \caption{Comparison of best NRMSE for each reconstruction method (across hyperparameter choices) between reconstructed images from 2.5\% real \FDG data and a standard OSEM reconstruction from 100\% count data. Error bars reflect a 95\% confidence interval across 10 different realizations of Poisson noisy data.}
    \label{fig:nrmse_real}
\end{figure}

Figure \ref{fig:nrmse_real} shows the minimum global NRMSE of each method when reconstructing from 2.5\% of full-count data, evaluated against full-count data. From this chart, we can see our proposed method marginally outperforms the Bowsher method for this dataset on quantitative reconstruction accuracy, while our proposed method strongly outperforms both of the other diffusion-model-based reconstruction algorithms.

\begin{figure*}
    \centering
    \includegraphics[width=1\textwidth]{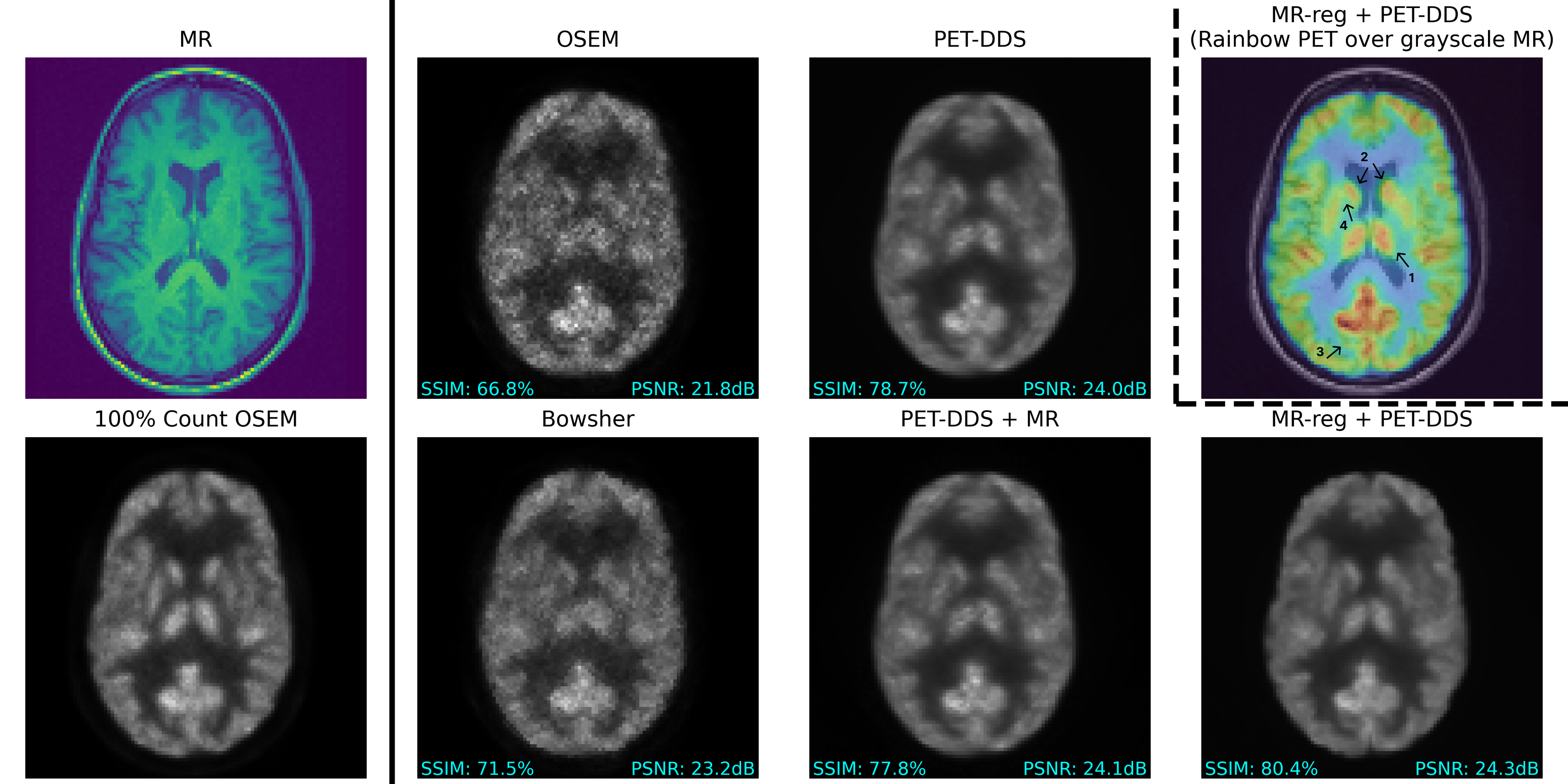}
    \caption{Example reconstructed transverse slices from real 2.5\% \FDG data, together with the corresponding T1 MRI and 100\% OSEM reconstruction. Hyperparameters for each method were chosen to maximize SSIM. Grayscale images are shown on the same linear scale. An image slice for MR-reg + PET-DDS is shown twice: once in grayscale for comparison with other methods (bottom-right), and additionally with rainbow-colored PET overlaid onto grayscale MR (top-right) to highlight areas of difference relative to other methods. Arrow 1: right thalamus. Arrow 2: caudate nucleus. Arrow 3: sulcal configurations. Arrow 4: the separation between the caudate and putamen.}
    \label{fig:images_real}
\end{figure*}

Figure \ref{fig:images_real} shows qualitative reconstructions from each reconstruction algorithm on real 2.5\% \FDG data. From this figure, we see that the diffusion-model-based approaches have suppressed noise throughout. Our proposed MR-reg + PET-DDS method has notably best preserved several brain structures, including: the right thalamus (arrow 1); the caudate nucleus (arrows 2); sulcal configurations (arrow 3); and, the separation between the caudate and putamen (arrow 4), which is resolved at least as well as the 100\% dose OSEM image.

\begin{figure}
    \centering
    \includegraphics[width=\linewidth]{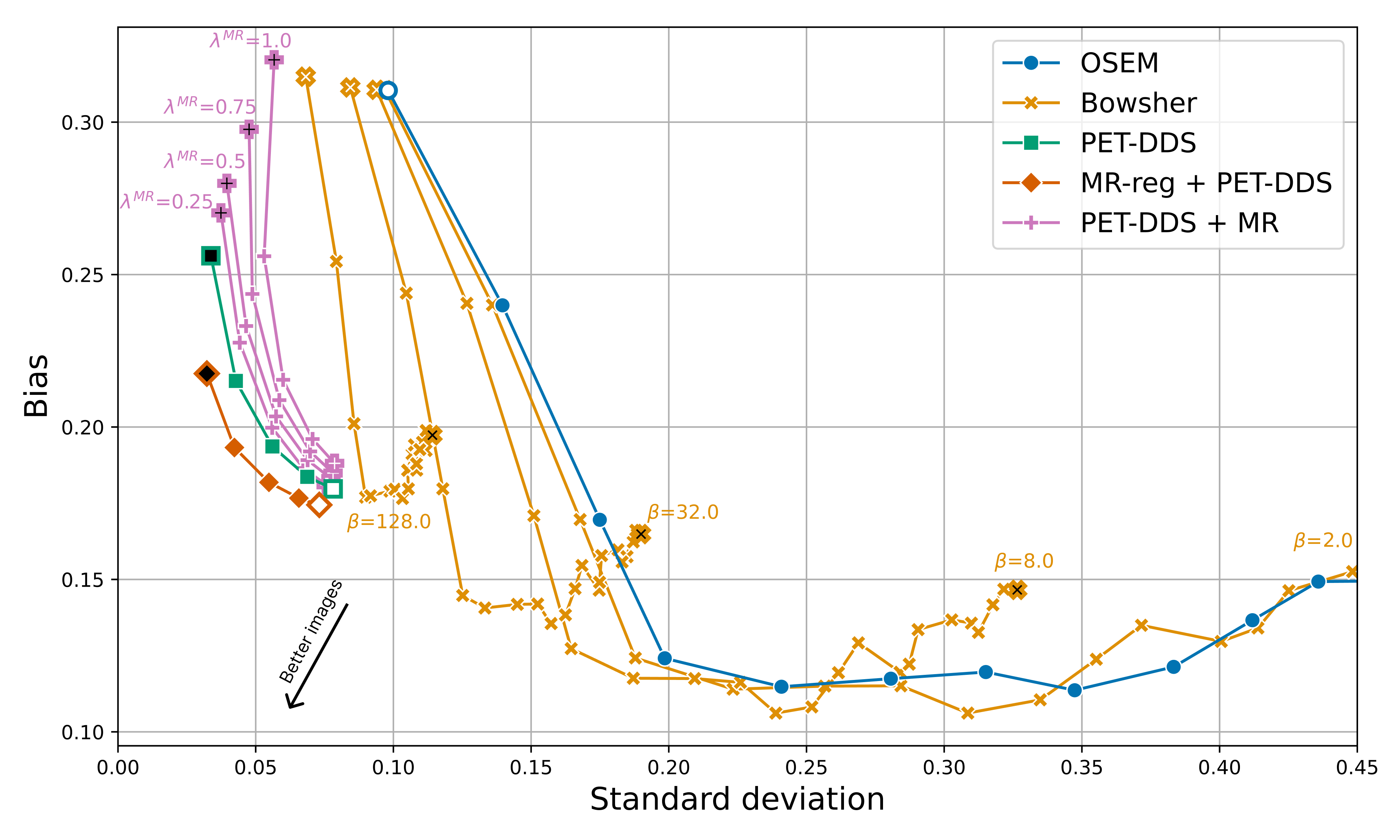}
    \caption{Bias-variance trade-off comparison between different reconstruction methods from 2.5\% real \FDG data. See Figure 5 for marker meanings.}
    \label{fig:bias_variance_real}
\end{figure}

To understand these results better, we performed a bias-variance trade-off assessment for each reconstruction algorithm, shown in Figure \ref{fig:bias_variance_real}. We see that the diffusion-model-based algorithms exhibit high bias, but low variance, whereas the conventional algorithms (OSEM and Bowsher) exhibit lower bias but higher variance. Bowsher can partially bridge these two regimes, at the cost of reconstructions with important artifacts (see Fig. \ref{fig:sim_images}). Of the diffusion-model-based approaches, our proposed approach MR-reg + PET-DDS achieves the lowest variance for a given bias (and vice versa), demonstrating its quantitative superiority.

\subsection{Computational efficiency}

Table \ref{tbl:timings} gives wall-clock timings for each approach. Training a registration network took up to 20.0 GB of GPU memory, while training a diffusion model took up to 19.7 GB of GPU memory. Performing reconstruction with the diffusion model and model-based data fidelity updates took up to 22.5 GB of GPU memory.

\begin{table}[t]
	\caption[]{Wall-clock timings recorded for each algorithm on reconstruction from 3D simulated FDG data.}
	\label{tbl:timings}
	\begin{center}
		\begin{sc}
            \input{timings}
		\end{sc}
	\end{center}
\end{table}

\section{Discussion}

Our work on simulated datasets reproduces and validates the work of Singh \textit{et al.} \cite{singh_score-based_2024} for PET-only reconstruction. For the real data case, we found that PET-DDS + MR did not offer a benefit over PET-DDS (see Figures \ref{fig:nrmse_real} and \ref{fig:bias_variance_real}). We hypothesize that this may be because learning from PET and MR images is a more complex problem than just learning from PET images, and that with insufficient training datasets, the benefits of MR information could not be fully realized. In the simulated case, the correspondence between the PET and MR images was much greater, which enabled the MR information to be better utilized.

In contrast, our proposed algorithm outperformed PET-DDS and PET-DDS + MR on both the simulated and real datasets. This suggests that using the MR datasets for pre-registration of PET data is better suited to the low-data regimes with fewer than $\sim 50$ paired images than training a diffusion model with MR data directly. Furthermore, training a diffusion model with MR data directly limits the quality of reconstruction to that of 100\% full-count OSEM (i.e. the training data); with the alternate choice of training data proposed, the same limit doesn't apply. 

As well as superior performance on the metrics considered, our reconstructions also have notably reduced background noise relative to other methods, which is important for the clinical utility of images \cite{hopson_clinical_2025}.

As presented, our method requires training a registration network and diffusion model for each reconstruction target, and so may be considered computationally intensive (see Table \ref{tbl:timings}). However, the registration time could be massively alleviated by training a general-purpose registration function on many MR volumes in advance. Furthermore, a patch-based approach to diffusion model training (using small patches) would further cut down the computational burden of using this method \cite{hu_patch-based_2024}, as would fine-tuning a pre-trained model.

Beyond improving efficiency, several avenues remain for further developing this methodology. More advanced MR registration algorithms, including those incorporating segmentation-based losses \cite{balakrishnan_voxelmorph_2019}, may yield higher-quality training data. The pseudo-PET data generated could be used with different score-based prior approaches such as PET-DPS \cite{singh_score-based_2024} or likelihood-scheduling for PET reconstruction \cite{webber_likelihood_2025}. For this study, we chose to use MR-MR registration rather than PET-PET registration, as it is most likely to be accurate due to the superior anatomical detail in MR relative to PET, and also allowed our registration approach to be both tracer and dose-invariant. Our initial results with PET-PET registration suggest that it is not suitable for ultra-low dose imaging; however, it remains to be seen whether results improve with higher dose levels. 

In addition to PET image reconstruction, the methodology for generating subject-specific PET may have further applications. In particular, when we include the random summing aspect of the methodology and register images pairwise, we can generate an infinite number of diverse PET-like images from very limited training data. This may have the potential to alleviate concerns (e.g. hallucinations, lack of interpretability of results) associated with sampling images from black-box generative models for some medical imaging use cases.

\section{Conclusion}

A novel principled method for subject-specific PET image synthesis was proposed, that synthesizes subject-specific PET images from a target MR image and a different subject's paired PET-MR images, without relying on black-box generative algorithms. Using this methodology to generate subject-specific training data enhanced the reconstruction accuracy of a state-of-the-art PET reconstruction algorithm (trained on image-only data). In particular, an improved trade-off between lesion detectability and reconstruction accuracy was observed for out-of-distribution lesions in simulations. We also showed that for real \FDG PET scans, our proposed approach can enhance the bias-variance characteristics of diffusion-model-based reconstruction and provide qualitatively more faithful reconstruction for low-dose PET-MR scans. It can be concluded that our method of data augmentation for training personalized diffusion models for reconstruction is a promising avenue for resolving out-of-distribution PET-MR mismatches, even when limited training data is available.

\section*{Acknowledgments}

All authors declare that they have no known conflicts of interest in terms of competing financial interests or personal relationships that could have an influence or are relevant to the work reported in this paper. Thanks to Dr Joel Dunn for his help anonymizing and pre-processing PET datasets.

%\bibliographystyle{ieeetr}
%\bibliography{bibliography}

\end{document}

%% file: timings.tex
\begin{tabularx}{\linewidth}{l|Y|Y|Y}
    \toprule
    \multirow{2}{*}{Method} & \multicolumn{3}{c}{Time} \\
    \cmidrule(lr){2-4}
    & Pre-train & Train & Recon\\
    \midrule
    OSEM & N/A & N/A & 4 s\\
    Bowsher & N/A & N/A & 38 s\\
    PET-DDS & N/A & 5 h 44m & 160 s\\
    PET-DDS+MR & N/A & 4h 23m & 252 s\\
    MR-reg + PET-DDS & 11h 52m & 4 h 27m & 160 s\\
    \bottomrule
\end{tabularx}

%% file: manuscript.bbl
\begin{thebibliography}{10}


\bibitem{shukla_positron_2006}

A.~K. Shukla and U.~Kumar, ``Positron emission tomography: An overview,'' {\em Journal of Medical Physics}, vol.~31, no.~1, pp.~13--21, 2006.

\newblock doi:10.4103/0971-6203.25665.


\bibitem{nievelstein_radiation_2012}

R.~A.~J. Nievelstein, H.~M.~E. Quarles~van Ufford, T.~C. Kwee, M.~B. Bierings, I.~Ludwig, F.~J.~A. Beek, J.~M.~H. de~Klerk, W.~P. T.~M. Mali, P.~W. de~Bruin, and J.~Geleijns, ``Radiation exposure and mortality risk from {CT} and {PET} imaging of patients with malignant lymphoma,'' {\em European Radiology}, vol.~22, no.~9, pp.~1946--1954, 2012.

\newblock doi:10.1007/s00330-012-2447-9.


\bibitem{schwaiger_petct_2005}

M.~Schwaiger, S.~Ziegler, and S.~G. Nekolla, ``{PET}/{CT}: Challenge for nuclear cardiology,'' {\em Journal of Nuclear Medicine}, vol.~46, no.~10, pp.~1664--1678, Oct. 2005.



\bibitem{tong_image_2010}

S.~Tong, A.~M. Alessio, and P.~E. Kinahan, ``Image reconstruction for {PET}/{CT} scanners: Past achievements and future challenges,'' {\em Imaging in Medicine}, vol.~2, no.~5, pp.~529--545, Oct. 2010.

\newblock doi:10.2217/iim.10.49.



\bibitem{bai_mr_2013}
B.~Bai, Q.~Li, and R.~M. Leahy, ``{MR} guided {PET} image reconstruction,'' {\em Seminars in Nuclear Medicine}, vol.~43, no.~1, pp.~30--44, Jan. 2013.

\newblock doi:10.1053/j.semnuclmed.2012.08.006.



\bibitem{corda-dincan_syn-net_2020}

G.~Corda-D'Incan, J.~A. Schnabel, and A.~J. Reader, ``Syn-{Net} for synergistic deep-learned {PET}-{MR} reconstruction,'' in {\em 2020 {IEEE} Nuclear Science Symposium and Medical Imaging Conference ({NSS}/{MIC})}, pp.~1--5, Oct. 2020.

\newblock doi:10.1109/NSS/MIC42677.2020.9508086.




\bibitem{chung_score-based_2022}
H.~Chung and J.~C. Ye, ``Score-based diffusion models for accelerated {MRI},'' {\em Medical Image Analysis}, vol.~80, p.~102479, Aug. 2022.

\newblock doi:10.1016/j.media.2022.102479.



\bibitem{chung_decomposed_2023}
H.~Chung, S.~Lee, and J.~C. Ye, ``Decomposed diffusion sampler for accelerating large-scale inverse problems,'' in {\em Proc. {ICLR}}, 2024.

\newblock (ICLR 2024 proceedings; also arXiv:2303.05754).



\bibitem{singh_score-based_2024}

I.~R. Singh {\em et~al.}, ``Score-based generative models for {PET} image reconstruction,'' {\em MELBA}, vol.~2, pp.~547--585, Jan. 2024.

\newblock doi:10.59275/j.melba.2024-5d51.



\bibitem{webber_likelihood_2025}

G.~Webber, Y.~Mizuno, O.~D. Howes, A.~Hammers, A.~P. King, and A.~J. Reader, ``Likelihood-scheduled score-based generative modeling for fully {3D} {PET} image reconstruction,'' {\em IEEE Transactions on Medical Imaging}, early access, 2025.

\newblock doi:10.1109/TMI.2025.3576483.



\bibitem{reader_deep_2021}

A.~J. Reader, G.~Corda, A.~Mehranian, C.~da~Costa-Luis, S.~Ellis, and J.~A. Schnabel, ``Deep learning for {PET} image reconstruction,'' {\em IEEE Transactions on Radiation and Plasma Medical Sciences}, vol.~5, no.~1, pp.~1--25, Jan. 2021.

\newblock doi:10.1109/TRPMS.2020.3014786.



\bibitem{shepp_maximum_1982}

L.~A. Shepp and Y.~Vardi, ``Maximum likelihood reconstruction for emission tomography,'' {\em IEEE Transactions on Medical Imaging}, vol.~1, no.~2, pp.~113--122, Oct. 1982.

\newblock doi:10.1109/TMI.1982.4307558.



\bibitem{levitan_maximum_1987}

E.~Levitan and G.~T. Herman, ``A maximum a posteriori probability expectation maximization algorithm for image reconstruction in emission tomography,'' {\em IEEE Transactions on Medical Imaging}, vol.~6, no.~3, pp.~185--192, Sept. 1987.

\newblock doi:10.1109/TMI.1987.4307826.



\bibitem{hudson_accelerated_1994}

H.~M. Hudson and R.~S. Larkin, ``Accelerated image reconstruction using ordered subsets of projection data,'' {\em IEEE Transactions on Medical Imaging}, vol.~13, no.~4, pp.~601--609, 1994.

\newblock doi:10.1109/42.363108.




\bibitem{de_pierro_fast_2001}
A.~De~Pierro and M.~Yamagishi, ``Fast {EM}-like methods for maximum ''a posteriori'' estimates in emission tomography,'' {\em IEEE Transactions on Medical Imaging}, vol.~20, no.~4, pp.~280--288, Apr. 2001.

\newblock doi:10.1109/42.921477.

\bibitem{bowsher_utilizing_2004}

J.~E. Bowsher, H.~Yuan, L.~W. Hedlund, T.~G. Turkington, G.~Akabani, A.~Badea, W.~C. Kurylo, C.~T. Wheeler, G.~P. Cofer, M.~W. Dewhirst, and G.~A. Johnson, ``Utilizing {MRI} information to estimate {F18}-{FDG} distributions in rat flank tumors,'' in {\em Proc. {IEEE} Nuclear Science Symposium and Medical Imaging Conference}, vol.~4, pp.~2488--2492, Oct. 2004.

\newblock doi:10.1109/NSSMIC.2004.1462760.



\bibitem{schramm_evaluation_2018}
G.~Schramm, M.~Holler, A.~Rezaei, K.~Vunckx, F.~Knoll, K.~Bredies, F.~Boada, and J.~Nuyts, ``Evaluation of parallel level sets and Bowsher’s method as segmentation-free anatomical priors for time-of-flight {PET} reconstruction,'' {\em IEEE Transactions on Medical Imaging}, vol.~37, no.~2, pp.~590--603, Feb. 2018.

\newblock doi:10.1109/TMI.2017.2767940.



\bibitem{ho_denoising_2020}
J.~Ho, A.~Jain, and P.~Abbeel, ``Denoising diffusion probabilistic models,'' in \emph{Advances in Neural Information Processing Systems 33 (NeurIPS 2020)}, H.~Larochelle, M.~Ranzato, R.~Hadsell, M.~F.~Balcan, and H.~Lin (eds.), Curran Associates, Inc., 2020, pp.~6840--6851.

\newblock doi:10.5555/3495724.3496298.

\bibitem{sohl-dickstein_deep_2015}
J.~Sohl-Dickstein, E.~Weiss, N.~Maheswaranathan, and S.~Ganguli, ``Deep unsupervised learning using nonequilibrium thermodynamics,'' in {\em Proceedings of the 32nd International Conference on Machine Learning (ICML)},
F.~Bach and D.~Blei, Eds., vol.~37 of {\em Proceedings of Machine Learning Research}, Lille, France, 2015, pp.~2256--2265.

\newblock doi:10.5555/3045118.3045358.

\bibitem{song_improved_2020}
Y.~Song and S.~Ermon,
``Improved techniques for training score-based generative models,'' in \emph{Advances in Neural Information Processing Systems 33 (NeurIPS 2020)}, H.~Larochelle, M.~Ranzato, R.~Hadsell, M.~F.~Balcan, and H.~Lin (eds.),
Curran Associates, Inc., 2020, pp.~12438--12448. 

\newblock doi:10.5555/3495724.3496767.


\bibitem{chung_solving_2023}
H.~Chung, D.~Ryu, M.~T. McCann, M.~L. Klasky, and J.~C. Ye,
``Solving {3D} inverse problems using pre-trained {2D} diffusion models,'' in \emph{Proceedings of the IEEE/CVF Conference on Computer Vision and Pattern Recognition (CVPR)}, Vancouver, BC, Canada, June 2023, pp.~22542--22551.

\newblock doi:10.1109/CVPR52729.2023.02159.




\bibitem{webber_diffusion_2024}

G.~Webber and A.~J. Reader, ``Diffusion models for medical image reconstruction,'' {\em {BJR}{\textbar}Artificial Intelligence}, p.~ubae013, Aug. 2024.

\newblock doi:10.1093/bjrai/ubae013.



\bibitem{ho_classifier-free_2022}
J.~Ho and T.~Salimans, ``Classifier-Free Diffusion Guidance,'' in \emph{Workshop on Deep Generative Models and Downstream Applications at NeurIPS 2021}, virtual, 2021.

\newblock doi:10.48550/arXiv.2207.12598.




\bibitem{balakrishnan_voxelmorph_2019}
G.~Balakrishnan, A.~Zhao, M.~R. Sabuncu, J.~V. Guttag, and A.~V. Dalca, ``{VoxelMorph}: A learning framework for deformable medical image registration,'' {\em IEEE Transactions on Medical Imaging}, vol.~38, no.~8, pp.~1788--1800, Aug. 2019.

\newblock doi:10.1109/TMI.2019.2897538.




\bibitem{bland_mr-guided_2018}
J.~Bland, A.~Mehranian, M.~A. Belzunce, S.~Ellis, C.~J. McGinnity, A.~Hammers, and A.~J. Reader,
``{MR}-guided kernel {EM} reconstruction for reduced dose {PET} imaging,'' {\em IEEE Transactions on Radiation and Plasma Medical Sciences}, vol.~2, no.~3, pp.~235--243, May 2018.

\newblock doi:10.1109/TRPMS.2017.2771490.



\bibitem{mehranian_model-based_2020}
A.~Mehranian and A.~J. Reader,
``Model-based deep learning {PET} image reconstruction using forward–backward splitting expectation–maximization,''
{\em IEEE Transactions on Radiation and Plasma Medical Sciences}, vol.~5, no.~1, pp.~54--64, June 2020.

\newblock doi:10.1109/TRPMS.2020.3004408.


\bibitem{xie_anatomically_2021}
Z.~Xie, T.~Li, X.~Zhang, W.~Qi, E.~Asma, and J.~Qi,
``Anatomically aided {PET} image reconstruction using deep neural networks,'' {\em Medical Physics}, vol.~48, no.~9, pp.~5244--5258, Sept. 2021.

\newblock doi:10.1002/mp.15051.



\bibitem{mehranian_pet_2017}
A.~Mehranian, M.~A. Belzunce, F.~Niccolini, M.~Politis, C.~Prieto, F.~Turkheimer, A.~Hammers, and A.~J. Reader, ``{PET} image reconstruction using multi-parametric anato-functional priors,'' {\em Physics in Medicine and Biology}, vol.~62, no.~15, p.~5975-6007, July 2017.

\newblock doi: 10.1088/1361-6560/aa7670.



\bibitem{costa-luis_micro-networks_2021}
C.~da~Costa-Luis and A.~J. Reader, ``Micro-networks for robust {MR}-guided low count {PET} imaging,'' {\em IEEE Transactions on Radiation and Plasma Medical Sciences}, vol.~5, no.~2, pp.~202--212, Mar. 2021.

\newblock doi:10.1109/TRPMS.2020.2986414.



\bibitem{gong_pet_2019}
K.~Gong, C.~Catana, J.~Qi, and Q.~Li,
``{PET} image reconstruction using deep image prior,''
{\em IEEE Transactions on Medical Imaging}, vol.~38, no.~7, pp.~1655--1665, July 2019.
\newblock doi:10.1109/TMI.2018.2888491.




\bibitem{burgos_subject-specific_2015}
N.~Burgos, M.~Jorge Cardoso, A.~F. Mendelson, J.~M. Schott, D.~Atkinson, S.~R. Arridge, B.~F. Hutton, and S.~Ourselin,
``Subject-specific models for the analysis of pathological {FDG} {PET} data,''
in {\em Medical Image Computing and Computer-Assisted Intervention -- MICCAI 2015},
N.~Navab, J.~Hornegger, W.~M. Wells, and A.~F. Frangi (eds.),
Lecture Notes in Computer Science, vol.~9350, Springer, Cham, 2015, pp.~651--658.

\newblock doi:10.1007/978-3-319-24571-3\_78.

\bibitem{webber_multi-subject_2024}
G.~Webber, Y.~Mizuno, O.~D. Howes, A.~Hammers, A.~P. King, and A.~J. Reader,
``Multi-subject image synthesis as a generative prior for single-subject {PET} image reconstruction,''
in {\em 2024 {IEEE} Nuclear Science Symposium (NSS), Medical Imaging Conference (MIC) and Room Temperature Semiconductor Detector Conference (RTSD)},
{IEEE}, Oct. 2024, pp.~1--2.

\newblock doi:10.1109/NSS/MIC/RTSD57108.2024.10657446.


\bibitem{modat_global_2014}
M.~Modat, D.~M. Cash, P.~Daga, G.~P. Winston, J.~S. Duncan, and S.~Ourselin, ``Global image registration using a symmetric block-matching approach,'' {\em Journal of Medical Imaging}, vol.~1, no.~2, p.~024003, Sept. 2014.

\newblock doi: 10.1117/1.JMI.1.2.024003.




\bibitem{dhariwal_diffusion_2021}
P.~Dhariwal and A.~Nichol, ``Diffusion models beat {GAN}s on image synthesis,'' in \emph{Advances in Neural Information Processing Systems 34 (NeurIPS 2021)}, M.~Ranzato, A.~Beygelzimer, Y.~Dauphin, P.~S.~Liang, and J.~W.~Vaughan (eds.), Curran Associates, Inc., 2021, pp.~8780--8794.

\newblock doi:10.5555/3540261.3540933.


\bibitem{ronneberger_u-net_2015}
O.~Ronneberger, P.~Fischer, and T.~Brox, ``{U}-{Net}: Convolutional networks for biomedical image segmentation,'' in {\em Medical Image Computing and Computer-Assisted Intervention -- MICCAI 2015}, N.~Navab, J.~Hornegger, W.~M.~Wells, and A.~F.~Frangi (eds.), Lecture Notes in Computer Science, vol.~9351, Springer, Cham, 2015, pp.~234--241.

\newblock doi:10.1007/978-3-319-24574-4\_28


\bibitem{vincent_connection_2011}
P.~Vincent,
``A connection between score matching and denoising autoencoders,''
{\em Neural Computation}, vol.~23, pp.~1661--1674, July 2011.

\newblock doi:10.1162/NECO\_a\_00142.



\bibitem{schramm_parallelprojopen-source_2024}
G.~Schramm and K.~Thielemans,
``PARALLELPROJ—an open-source framework for fast calculation of projections in tomography,''
{\em Frontiers in Nuclear Medicine}, vol.~3, Jan. 2024.

\newblock doi:10.3389/fnume.2023.1324562.



\bibitem{reader_bootstrap-optimised_2020}
A.~J. Reader and S.~Ellis,
``Bootstrap-optimised regularised image reconstruction for emission tomography,''
{\em IEEE Transactions on Medical Imaging}, vol.~39, no.~6, pp.~2163--2175, June 2020.

\newblock doi: 10.1109/TMI.2019.2956878.



\bibitem{hopson_clinical_2025}
J.~B. Hopson, S.~Ellis, A.~Flaus, C.~J. McGinnity, R.~Neji, A.~J. Reader, and A. Hammers,
``Clinical and deep-learned evaluation of {MR}-guided self-supervised {PET} reconstruction,''
{\em IEEE Transactions on Radiation and Plasma Medical Sciences}, vol.~9, no.~3, pp.~337--346, Mar. 2025.

\newblock doi:10.1109/TRPMS.2024.3496779.


\bibitem{hu_patch-based_2024}
J.~Hu, B.~Song, J.~A. Fessler, and L.~Shen,
``Patch-based diffusion models beat whole-image models for mismatched distribution inverse problems,''
arXiv:2410.11730, Oct. 2024.

\newblock doi:10.48550/arXiv.2410.11730.



\end{thebibliography}
